\documentclass[12pt,a4paper]{article}
\usepackage{amsmath,caption}
\usepackage[top=1.2in, bottom=1.2in, left=1.2in, right=1.2in]{geometry}

\DeclareCaptionStyle{italic}[justification=centering]{labelfont={bf},font ={small},textfont={it},labelsep=colon}
\usepackage{graphicx,psfrag,epsf}
\usepackage{enumerate, algorithm, algorithmic}
\usepackage{natbib}
\usepackage{url} 
\usepackage{booktabs, subfig, bm, paralist,mathpazo,tikz,todonotes,longtable,microtype,dsfont,rotating} 
\usepackage[pdftex,colorlinks=true]{hyperref}
\definecolor{darkblue}{rgb}{0,0,.6}
\hypersetup{citecolor=darkblue,linkcolor=darkblue,urlcolor=darkblue}

\newcommand{\blind}{0}

\graphicspath{{plots/}}
\DeclareMathOperator*{\argmin}{\arg\!\min}
\newsavebox\CBox
\def\textBF#1{\sbox\CBox{#1}\resizebox{\wd\CBox}{\ht\CBox}{\textbf{#1}}}

\definecolor{a0}{rgb}{0.0, 0.5, 0.0}
\definecolor{bistre}{rgb}{0.24, 0.17, 0.12}
\definecolor{amethyst}{rgb}{0.6, 0.4, 0.8}
\definecolor{blue-violet}{rgb}{0.54, 0.17, 0.89}
\definecolor{Rcolor}{RGB}{150,160,190}
\definecolor{blush}{rgb}{0.87, 0.36, 0.51}
\definecolor{brightturquoise}{rgb}{0.03, 0.91, 0.87}
\definecolor{burntorange}{rgb}{0.8, 0.33, 0.0}
\date{\today}
\AtBeginDocument{}
\begin{document}

\def\spacingset#1{\renewcommand{\baselinestretch}
{#1}\small\normalsize} \spacingset{1}

\if0\blind
{
  \title{\bf Forecasting age distribution of death counts: \mbox{An application to annuity pricing}}
  \author{Han Lin Shang\thanks{Correspondence to: Han Lin Shang, Research School of Finance, Actuarial Studies and Statistics, Level 4, Building 26C, Kingsley Street, Australian National University, Kingsley Street, Acton, Canberra, ACT 2601, Australia; Telephone: +61(2) 612 50535; Fax: +61(2) 612 50087; Email: hanlin.shang@anu.edu.au}
  \hspace{.2cm}\\
    Research School of Finance, Actuarial Studies and Statistics \\
    Australian National University \\
    \\
    Steven Haberman \\
    Cass Business School \\ 
    City, University of London}
  \maketitle
  \thispagestyle{empty}
} \fi

\if1\blind
{
  \bigskip
  \bigskip
  \bigskip
  \begin{center}
    {\LARGE\bf Forecasting age distribution of death counts: \mbox{An application to annuity pricing}}
\end{center}
  \medskip
} \fi

\bigskip

\begin{abstract}

We consider a compositional data analysis approach to forecasting the age distribution of death counts. Using the age-specific period life-table death counts in Australia obtained from the Human Mortality Database, the compositional data analysis approach produces more accurate one- to 20-step-ahead point and interval forecasts than Lee-Carter method, Hyndman-Ullah method, and two na\"{i}ve random walk methods. The improved forecast accuracy of period life-table death counts is of great interest to demographers for estimating survival probabilities and life expectancy, and to actuaries for determining temporary annuity prices for various ages and maturities. Although we focus on temporary annuity prices, we consider long-term contracts which make the annuity almost lifetime, in particular when the age at entry is sufficiently high.

\end{abstract}

\\

\newpage
\spacingset{1.48}

\section{Introduction}

Actuaries have produced forecasts of mortality since the beginning of the 20th century, in response to the adverse financial effects of mortality improvements over time on life annuities and pensions \citep{Pollard87}. Thus, projected mortality tables for annuitants was one of the topics discussed at the 5\textsuperscript{th} International Congress of Actuaries, Berlin in 1906 \citep[][]{CW35}. Several authors have proposed new approaches for forecasting age-specific central mortality rates using statistical models \citep[see][for reviews]{Booth06, BT08, CBD08, SBH11}. Instead of modelling central mortality rates, we consider a compositional data analysis (CoDa) approach for modelling and forecasting the age-specific numbers of deaths in period life tables. Both central mortality rates or life-table death counts can be derived from the other based on standard life-table relations (for detail on the life table and its indicators, see \cite{PHG01}, Chapter 3, or \cite{DHW09}, Chapters 2-3). By using the life-table death distribution, we could model and forecast a redistribution of the density of life-table deaths, where deaths at younger ages are shifted towards older ages. Alternatively, we may consider a cohort life table which depicts the life history of a specific group of individuals but is dependent on projected mortality rates for those cohorts born more recently. Instead, we choose to study the period life table which represents the mortality conditions in a period of time \citep[see also][]{Oeppen08, BC17}.

In the field of demography, \cite{Oeppen08} and \cite{BC17} have put forward a principal component approach to forecast life-table death counts within a CoDa framework by considering age-specific life-table death count $(d_x)$ as compositional data. As with compositional data, the data are constrained to vary between two limits (e.g., 0 and a constant upper bound), which conditions their covariance structure. This feature can represent great advantages in a forecasting context \citep{Lee98}. Thus, \cite{Oeppen08} demonstrated that using CoDa to forecast age-specific mortality does not lead to more pessimistic results than forecasting age-specific mortality \citep[see, e.g.,][]{Wilmoth95}. Apart from providing an informative description of the mortality experience of a population, the age-at-death distribution yields readily available information on the ``central longevity indicators" \citep[e.g., mean, median and modal age at death, see][]{CRT+05, Canudas-Romo10} as well as lifespan variability \citep[e.g.,][]{Robine01, VZV11, VC13, VMM14, AV18}.

Compositional data arise in many other scientific fields, such as geology (geochemical elements), economics (income/expenditure distribution), medicine (body composition), food industry (food composition), chemistry (chemical composition), agriculture (nutrient balance bionomics), environmental sciences (soil contamination), ecology (abundance of different species), and demography (life-table death counts). In the field of statistics, \cite{SCG+15} use CoDa to study the concentration of chemical elements in sediment or rock samples. \cite{SW17} applied CoDa to analyse total weekly expenditure on food and housing costs for households in a chosen set of domains. \cite{Delicado11} and \cite{KMP+18} use CoDa to analyse density functions and implement dimension-reduction techniques on the constrained compositional data space.

Compositional data are defined as a random vector of $K$ positive components $\bm{D} = [d_1,\dots,d_K]$ with strictly positive values whose sum is a given constant, set typically equal to 1 (portions), 100 (percentage) and $10^6$ for parts per million (ppm) in geochemical trace element compositions \citep[][p. 1]{Aitchison86}. The sample space of compositional data is thus the simplex
\begin{equation*}
\mathcal{S}^K = \left\{\bm{D} = \left(d_1, \dots, d_K\right)^{\top}, \quad d_x>0, \quad \sum^K_{x=1}d_x = c\right\},
\end{equation*}
where $\mathcal{S}$ denotes a simplex, $c$ is a fixed constant, $^{\top}$ denotes vector transpose, and the simplex sample space is a $K-1$ dimensional subset of real-valued space $R^{K-1}$.

Compositional data are subject to a sum constraint, which in turn imposes unpleasant constraints upon the variance-covariance structure of the raw data. The standard approach involves breaking the sum constraint using a transformation of the raw data to remove the constraint, before applying conventional statistical techniques to the transformed data. Among all possible transformations, the family of log-ratio transformations is commonly used. This family includes the additive log-ratio, the multiple log-ratio, the centred log-ratio transformations \citep{AS80, Aitchison82, Aitchison86}, and the isometric log-ratio transformation \citep{EPM+03}. 

The contributions of this paper are threefold: First, as the CoDa framework of \cite{Oeppen08} is an adaptation of the Lee-Carter model to compositional data, our work could be seen as an adaptation of \citeauthor{HU07}'s \citeyearpar{HU07} to a CoDa framework. Second, we apply the CoDa method of \cite{Oeppen08} to model and forecast the age distribution of life-table death counts, from which we obtain age-specific survival probabilities, and we determine immediate temporary annuity prices. Third, we propose a nonparametric bootstrap method for constructing prediction intervals for the future age distribution of life-table death counts.

Using the Australian age- and sex-specific life-table death counts from 1921 to 2014, we evaluate and compare the one- to 20-step-ahead point forecast accuracy and interval forecast accuracy among the CoDa, Hyndman-Ullah (HU) method, Lee-Carter (LC) method and two na\"{i}ve random walk methods. To evaluate point forecast accuracy, we use the mean absolute percentage error (MAPE). To assess interval forecast accuracy, we utilise the interval score of \cite{GR07} and \cite{GK14}, see Section~\ref{sec:forecast_error} for details. Regarding both point forecast accuracy and interval forecast accuracy, the CoDa method performs the best overall among the three methods which we have considered. The improved forecast accuracy of life-table death counts is of great importance to actuaries for determining remaining life expectancies and pricing temporary annuities for various ages and maturities. 

The remainder of this paper is organised as follows: Section~\ref{sec:2} describes the data set, which is Australian age- and sex-specific life-table death counts from 1921 to 2014. Section~\ref{sec:3} introduces the CoDa method for producing the point and interval forecasts of the age distribution of life-table death counts. Section~\ref{sec:model_fit} studies the goodness-of-fit of the CoDa method and provides an example for generating point and interval forecasts. Using the MAPE and mean interval score in Section~\ref{sec:4}, we evaluate and compare the point and interval forecast accuracies among the methods considered. Section~\ref{sec:6} applies the CoDa method to estimate the single-premium temporary immediate annuity prices for various ages and maturities for a \textit{female} policyholder residing in Australia. Conclusions are presented in Section~\ref{sec:7}, along with some reflections on how the methods presented here can be extended.

\section{Australian age- and sex-specific life-table death counts}\label{sec:2}

We consider Australian age- and sex-specific life-table death counts from 1921 to 2014, obtained from the \cite{HMD17}. Although we use all data that include the first and second World War periods, a feature of our proposed method is to use an automatic algorithm of \cite{HK08} to select the optimal data-driven parameters in exponential smoothing forecasting method. As with exponential smoothing forecasting method, it assigns more weights to the recent data than distant past data. In practice, it is likely that only a few of the most distant past data may influence our forecasts, regardless of the starting point.

We study life-table death counts, where the life-table radix (i.e., a population experiencing 100,000 births annually) is fixed at 100,000 at age 0 for each year. For the life-table death counts, there are 111 ages, and these are age 0, 1, $\cdots$, 109, 110+. Due to rounding, there are zero counts for age 110+ at some years. To rectify this problem, we prefer to use the probability of dying (i.e., $q_x$) and the life-table radix to recalculate our estimated death counts (up to 6 decimal places). In doing so, we obtain more detailed death counts than the ones reported in the \cite{HMD17}.

\begin{figure}[!htbp]
\centering
{\includegraphics[width=8.2cm]{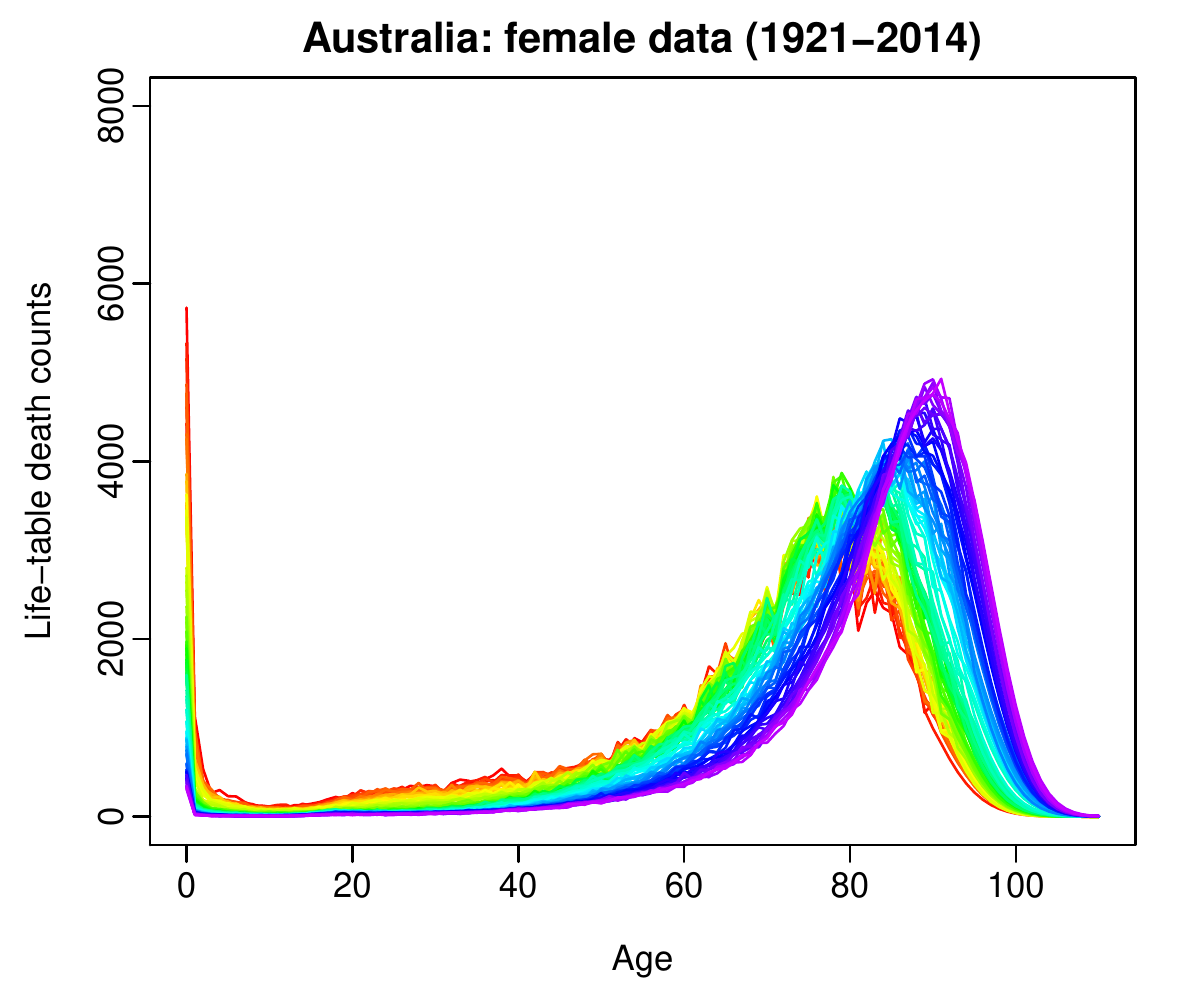}}
\qquad
{\includegraphics[width=8.2cm]{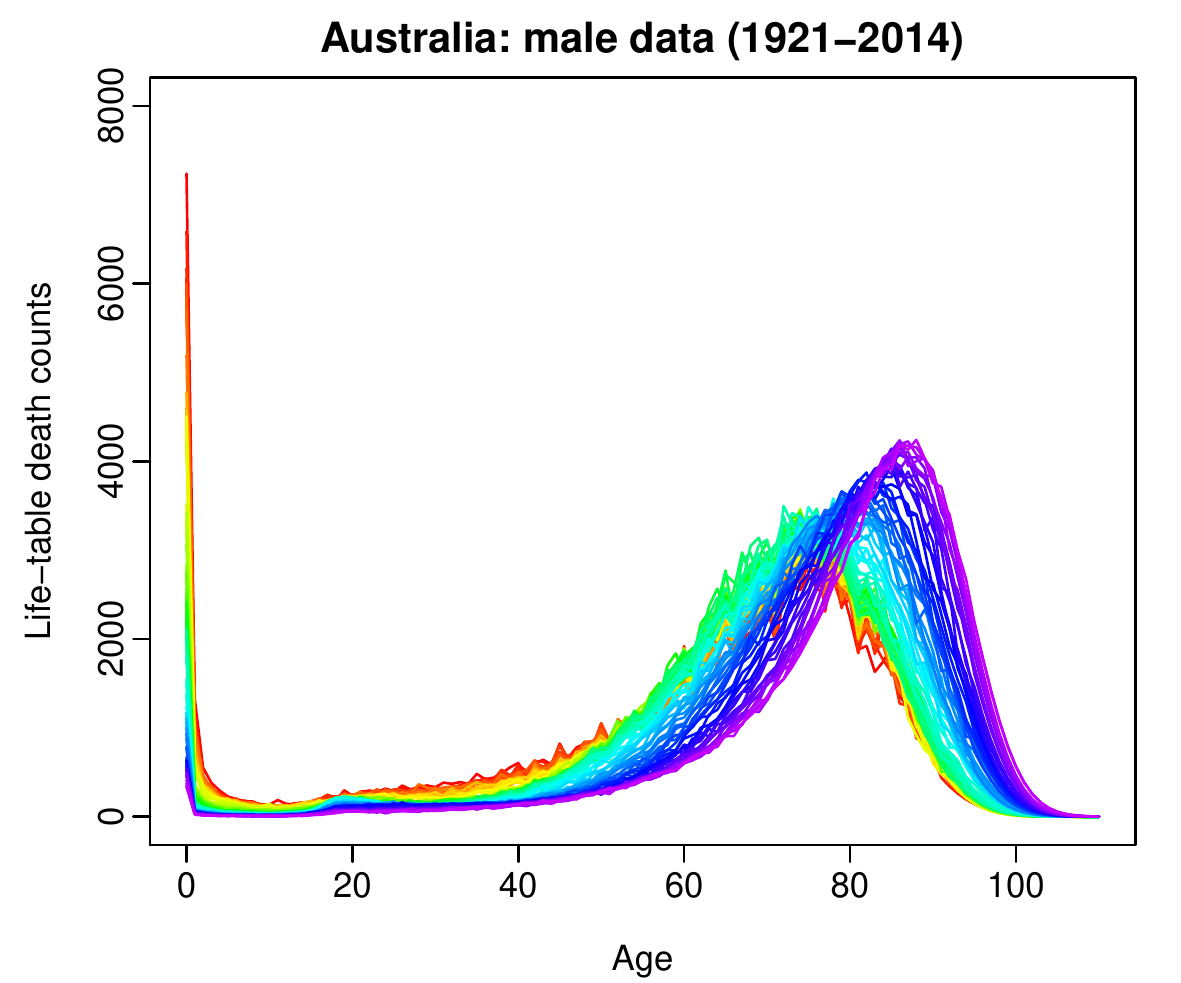}}
\caption{Rainbow plots of age-specific life-table death count from 1921 to 2014 in a single-year group. The oldest years are shown in red, with the most recent years in violet. Curves are ordered chronologically according to the colours of the rainbow.}\label{fig:1}
\end{figure}

To understand the principal features of the data, Figure~\ref{fig:1} presents rainbow plots of the female and male age-specific life-table death counts in Australia from 1921 to 2014 in a single-year group. The time ordering of the curves follows the colour order of a rainbow, where data from the distant past are shown in red and the more recent data are shown in purple \citep[see][for other examples]{HS10}. Both figures demonstrate a decreasing trend in infant death counts and a typical negatively-skewed distribution for the life-table death counts, where the peaks shift to higher ages for both females and males. This shift is a primary source of the longevity risk, which is a major issue for insurers and pension funds, especially in the selling and risk management of annuity products \citep[see][for a discussion]{DDG07}. Moreover, the spread of the distribution indicates lifespan variability. A decrease in variability over time can be observed directly and can be measured, for example with the interquartile range of life-table ages at death or the Gini coefficient \citep[for comprehensive reviews, see][]{WH99, VC13, DCH+17}. Age-at-death distributions thus provide critical insights on longevity and lifespan variability that cannot be grasped directly from either mortality rates or the survival function.

\section{Forecasting method}\label{sec:3}

\subsection{Compositional data analysis (CoDa)}\label{sec:3.1}

In the field of demography, \cite{Oeppen08} and \cite{BC17} have laid the foundations by presenting a modelling and forecasting framework. Following \cite{Oeppen08} and \cite{BC17}, a CoDa method can be summarised as follows:
\begin{enumerate}[1)]
\item We begin from a data matrix $\bm{D}$ of size $n\times K$ of the life-table death counts $(d_{t,x})$ with $n$ rows representing the number of years and $K$ columns representing the age $x$. The sum of each row adds up to the life-table radix, such as 100,000. Since we are working with life-table death counts, it is not necessary to have population-at-risk estimates.
\item We compute the geometric mean at each age, given by
\begin{equation}
\alpha_x = \exp^{\frac{1}{n}\sum^n_{t=1}\ln (d_{t,x})}, \qquad x = 1,\dots, K,\quad t=1,\dots,n.\label{eq:geometric_mean}
\end{equation}
For a given year $t$, we divide $(d_{t,1},\dots,d_{t,K})$ by the corresponding geometric means $(\alpha_1,\dots,\alpha_K)$, then standardise all elements so as to sum up to unity. As with compositional data, it is more important to know a relative proportion of each component (i.e., age) than the sum of all components. Although the life-table radix is 100,000 customarily, we model and forecast relative proportions, which are then multiplied by 100,000 to obtain forecast life-table death counts. Via standardisation, this is expressed as
\begin{equation*}
f_{t,x} = \frac{d_{t,x}/\alpha_x}{\sum^K_{x=1}d_{t,x}/\alpha_x},
\end{equation*}
where $f_{t,x}$ denotes de-centred data. 
\item Log-ratio transformation: \cite{Aitchison82, Aitchison86} showed that compositional data are represented in a restricted space where the components can only vary between 0 and a positive constant. Therefore, \cite{Aitchison82, Aitchison86} proposed a log-ratio transformation to transform the data into a real-valued space. We apply the centred log-ratio transformation, given by
\begin{equation*}
z_{t,x} = \ln\left(\frac{f_{t,x}}{g_t}\right),
\end{equation*}
where $g_t$ denotes the geometric mean over the age at time $t$, given by
\begin{equation*}
g_t = \exp^{\frac{1}{K}\sum^K_{x=1}\ln (f_{t,x})}.
\end{equation*}
The transformed data matrix is denoted as $\bm{Z}$ with elements $z_{t,x}\in R$, where $R$ denotes real-valued space.
\item Principal component analysis: Principal component analysis is then applied to the matrix $\bm{Z}$, to obtain the estimated principal components and their scores,
\begin{equation}
z_{t,x} = \sum_{\ell=1}^{\min(n,K)}\widehat{\beta}_{t,\ell}\widehat{\phi}_{\ell,x} = \sum^L_{\ell=1}\widehat{\beta}_{t,\ell}\widehat{\phi}_{\ell,x}+\widehat{\varpi}_{t,x}, \label{eq:ncomp}
\end{equation}
where $\widehat{\varpi}_{t,x}$ denotes model residual term for age $x$ in year $t$, $\{\widehat{\phi}_{1,x},\dots,\widehat{\phi}_{L,x}\}$ represents the first $L$ sets of estimated principal components, $\{\widehat{\beta}_{t,1},\dots,\widehat{\beta}_{t,L}\}$ represents the first $L$ sets of estimated principal component scores for time $t$, and $L$ denotes the number of retained principal components.
\item Forecast of principal component scores: Via a univariate time series forecasting method, such as exponential smoothing (ETS), we obtain the $h$-step-ahead forecast of the $\ell^{\text{th}}$ principal component score $\widehat{\beta}_{n+h, \ell}$, where $h$ denotes forecast horizon \citep[see also][]{HU07}. We utilise an automatic algorithm developed by \cite{HK08} to determine the optimal ETS model based on the corrected Akaike Information Criterion \citep{HT93}. Conditioning on the estimated principal components and observed data, the forecast of $z_{n+h,x}$ can be obtained by
\begin{equation}
\widehat{z}_{n+h|n, x} = \sum^{L}_{\ell=1}\widehat{\beta}_{n+h|n,\ell}\widehat{\phi}_{\ell,x}. \label{eq:2}
\end{equation}
In~\eqref{eq:2}, the principal component scores can be modelled and forecasted by an ETS method. To select the optimal ETS parameter, we use the automatic search algorithm of \cite{HK08} by the smallest corrected Akaike information criterion. In Table~\ref{tab:CoDa_results}, we also consider three other forecasting methods, namely autoregressive integrated moving average (ARIMA), random walk with drift (RWD) and random walk without drift (RW).
\item Transform back to the compositional data: we take the inverse centred log-ratio transformation, given by
\begin{equation*}
\widehat{\bm{f}}_{n+h|n} = \left[\frac{\exp^{\widehat{z}_{n+h|n, 1}}}{\sum^K_{x=1}\exp^{\widehat{z}_{n+h|n, x}}}, \frac{\exp^{\widehat{z}_{n+h|n, 2}}}{\sum^K_{x=1}\exp^{\widehat{z}_{n+h|n, x}}}, \dots,\frac{\exp^{\widehat{z}_{n+h|n, K}}}{\sum^K_{x=1}\exp^{\widehat{z}_{n+h|n, x}}}\right],
\end{equation*}
where $\widehat{z}_{n+h|n, x}$ denotes the forecasts in~\eqref{eq:2}.
\item Finally, we add back the geometric means, to obtain the forecasts of the life-table death matrix $d_{n+h}$,
\begin{align*}
\widehat{\bm{d}}_{n+h|n} &= \left[\frac{\widehat{f}_{n+h|n, 1}\times \alpha_1}{\sum^K_{x=1}\widehat{f}_{n+h|n, x}\times \alpha_x}, \frac{\widehat{f}_{n+h|n,2}\times \alpha_2}{\sum^K_{x=1}\widehat{f}_{n+h|n,x}\times \alpha_x},\dots,\frac{\widehat{f}_{n+h|n,K}\times \alpha_K}{\sum^K_{x=1}\widehat{f}_{n+h|n,x}\times \alpha_x}\right],
\end{align*}
where $\alpha_x$ is the age-specific geometric mean given in~\eqref{eq:geometric_mean}.
\end{enumerate}

To determine the number of components $L$ in~\eqref{eq:ncomp} and~\eqref{eq:2}, we consider a criterion known as cumulative percentage of variance (CPV), i.e., define the value of $L$ as the minimum number of components that reaches a certain level of the proportion of total variance explained by the $L$ leading components such that
\begin{equation*}
L = \argmin_{L: L\geq 1} \left\{\sum^L_{\ell=1}\widehat{\lambda}_{\ell}\bigg/\sum^{n}_{\ell=1}\widehat{\lambda}_{\ell}\mathds{1}_{\left\{\widehat{\lambda}_{\ell}>0\right\}} \geq \delta\right\},
\end{equation*}
where $\delta = 85\%$ \citep[see also][p.41]{HK12} and $\mathds{1}_{\{\cdot\}}$ denotes the binary indicator function which excludes possible zero eigenvalues. For the Australian female and male data, the chosen number of components $L=1$ and $L=2$, respectively. 

We highlight some similarities and differences between \citeauthor{Oeppen08}'s \citeyearpar{Oeppen08} approach and \citeauthor{LC92}'s \citeyearpar{LC92} approach. From Steps 3 to 5, \citeauthor{Oeppen08}'s \citeyearpar{Oeppen08} approach uses the \citeauthor{LC92}'s \citeyearpar{LC92} approach to model and forecast the log-ratio of life-table death counts. The difference is that the \citeauthor{Oeppen08}'s \citeyearpar{Oeppen08} approach works with life-table death counts in a constrained space, whereas the \citeauthor{LC92}'s \citeyearpar{LC92} approach works with real-valued log mortality rates.

When the number of components $L=1$, our proposed CoDa method corresponds to the one presented by \cite{Oeppen08} and \cite{BC17}. However, we allow the possibility of using more than one pair of principal component and principal component scores in Steps 4 and 5. As a sensitivity analysis, we also consider setting the number of components to be $L=6$ \citep[see also][]{HB08}. From this aspect, our proposal shares some similarity with the \citeauthor{HU07}'s \citeyearpar{HU07} approach. The difference is that our proposal works with life-table death counts in a constrained space, whereas the \citeauthor{HU07}'s \citeyearpar{HU07} approach works with real-valued log mortality rates.

\subsection{Construction of prediction interval for the CoDa}\label{sec:CoDa_PI}

Prediction intervals are a valuable tool for assessing the probabilistic uncertainty associated with point forecasts. The forecast uncertainty stems from both systematic deviations (e.g., due to parameter or model uncertainty) and random fluctuations (e.g., due to model error term). As was emphasised by \cite{Chatfield93, Chatfield00}, it is essential to provide interval forecasts as well as point forecasts to
\begin{enumerate}[1)]
\item assess future uncertainty levels;
\item enable different strategies to be planned for a range of possible outcomes indicated by the interval forecasts;
\item compare forecasts from different methods more thoroughly; and
\item explore different scenarios based on various assumptions.
\end{enumerate}

\subsubsection{Proposed bootstrap method}\label{sec:3.2.1}

We consider two sources of uncertainty: truncation errors in the principal component decomposition and forecast errors in the projected principal component scores. Since principal component scores are regarded as surrogates of the original functional time series, these principal component scores capture the temporal dependence structure inherited in the original functional time series \citep[see also][]{SG15, Paparoditis17, Shang18}. By adequately bootstrapping the forecast principal component scores, we can generate a set of bootstrapped $\bm{Z}^*$, conditional on the estimated mean function and estimated functional principal components from the observed $Z$ in~\eqref{eq:2}.

Using a univariate time series model, we can obtain multi-step-ahead forecasts for the principal component scores, $\{\widehat{\beta}_{1,\ell},\dots,\widehat{\beta}_{n,\ell}\}$ for $l=1,\dots,L$. Let the $h$-step-ahead forecast errors be given by $\widehat{\vartheta}_{t,h,\ell} = \widehat{\beta}_{t,\ell}-\widehat{\beta}_{t|t-h,\ell}$ for $t=h+1,\dots,n$. These can then be sampled with replacement to give a bootstrap sample of $\beta_{n+h,\ell}$:
\begin{equation*}
\widehat{\beta}_{n+h|n,\ell}^{(b)} = \widehat{\beta}_{n+h|n,\ell}+\widehat{\vartheta}_{\ast, h, \ell}^{(b)}, \qquad b=1,\dots,B,
\end{equation*}
where $B=1,000$ symbolises the number of bootstrap replications and $\widehat{\vartheta}_{\ast, h, \ell}^{(b)}$ are sampled with replacement from $\{\widehat{\vartheta}_{t, h, \ell}\}$.

Assuming the first $L$ principal components approximate the data $\bm{Z}$ relatively well, the model residual should contribute nothing but random noise. Consequently, we can bootstrap the model fit errors in~\eqref{eq:ncomp} by sampling with replacement from the model residual term $\{\widehat{\varpi}_{1,x},\dots,\widehat{\varpi}_{n,x}\}$ for a given age $x$. 

Adding all two components of variability, we obtain $B$ variants for $z_{n+h,x}$,
\begin{equation*}
\widehat{z}_{n+h|n,x}^{(b)} = \sum^L_{\ell=1}\widehat{\beta}^{(b)}_{n+h|n,\ell}\widehat{\phi}_{\ell,x}+\widehat{\varpi}_{n+h,x}^{(b)},
\end{equation*}
where $\widehat{\beta}^{(b)}_{n+h,\ell}$ denotes the forecast of the bootstrapped principal component scores. The construction of the bootstrap samples $\{\widehat{z}^{(1)}_{n+h|n, x}, \dots, \widehat{z}^{(B)}_{n+h|n,x}\}$ is in the same spirit as the construction of the bootstrap samples for the LC method in~\eqref{eq:LC_PI} and for the HU method in \cite{HS10}.

With the bootstrapped $\widehat{z}_{n+h|n,x}^{(b)}$, we follow steps 6) and 7) in Section~\ref{sec:3.1}, in order to obtain the bootstrap forecasts of $d_{n+h,x}$. At the $100(1-\gamma)\%$ nominal coverage probability, the pointwise prediction intervals are obtained by taking $\gamma/2$ and $1-\gamma/2$ quantiles based on $\{\widehat{d}_{n+h|n,x}^{(1)},\dots,\widehat{d}_{n+h|n,x}^{(B)}\}$.

\subsubsection{Existing bootstrap method}\label{sec:3.2.2}

The construction of the prediction interval for CoDa has also been considered by \cite{BC17}. After applying the centred log-ratio transformation, \cite{BC17} fit a nonparametric model to estimate the regression mean and obtain the residual matrix, from which bootstrap residual matrices can be obtained by randomly sampling with replacement from the original residual matrix. With the bootstrapped residuals, they then add them to the estimated regression mean to form the bootstrapped data samples. With each replication of the bootstrapped data samples, one could then re-apply singular value decomposition to obtain the first set of principal component and its associated scores. With the bootstrapped first set of principal component scores, one could then extrapolate them to the future using a univariate time-series forecasting method. By multiplying the bootstrapped forecast of the principal component scores with the bootstrap principal component, bootstrap forecasts of life-table death counts can be obtained via back-transformation.

In Table~\ref{tab:CoDa_results}, we compare the interval forecast accuracy, as measured by the mean interval score, between the two ways of constructing prediction intervals (as described in Section~\ref{sec:3.2.1} and~\ref{sec:3.2.2}.

\subsection{Other forecasting methods}

\subsubsection{Lee-Carter (LC) method}

As a comparison, we revisit \citeauthor{LC92}'s \citeyearpar{LC92} method. To stabilise the higher variance associated with mortality at advanced old ages, it is necessary to transform the raw data by taking the natural logarithm. For those missing death counts at advanced old ages, we use a simple linear interpolation method to approximate the missing values. We denote by $m_{t,x}$ the observed mortality rate at year $t$ at age $x$ calculated as the number of deaths in the calendar year $t$ at age $x$, divided by the corresponding mid-year population aged $x$. The model structure is given by
\begin{equation}
\ln (m_{t,x}) = a_x + b_x\kappa_t + \epsilon_{t,x},\label{eq:LC}
\end{equation}
where $a_x$ denotes the age pattern of the log mortality rates averaged across years; $b_x$ denotes the first principal component reflecting relative change in the log mortality rate at each age; $\kappa_t$ is the first set of principal component scores by year $t$ and measures the general level of the log mortality rates; and $\epsilon_{t,x}$ denotes the residual at year $t$ and age $x$.

The LC model in~\eqref{eq:LC} is over-parametrised in the sense that the model structure is invariant under the following transformations:
\begin{align*}
  \{a_x, b_x, \kappa_t\} &\mapsto \{a_x, b_x/c, c\kappa_t\}, \\
  \{a_x, b_x, \kappa_t\} &\mapsto \{a_x-cb_x, b_x, \kappa_t+c\}.
\end{align*}
To ensure the model's identifiability, \citet{LC92} imposed two constraints, given as:
\begin{equation*}
  \sum^n_{t=1}\kappa_t=0, \qquad \sum^{x_p}_{x=x_1}b_x=1,
\end{equation*}
where $n$ is the number of years, and $p$ is the number of ages in the observed data set.

Also, the LC method adjusts $\kappa_t$ by refitting to the total number of deaths. This adjustment gives more weight to high rates, thus roughly counterbalancing the effect of using a log transformation of the mortality rates. The adjusted $k_{t}$ is then extrapolated using RW models. \citet{LC92} used an RWD model, which can be expressed as:
\begin{equation*}
  \kappa_{t}=\kappa_{t-1}+d+e_{t},
\end{equation*}
where $d$ is known as the drift parameter and measures the average annual change in the series, and $e_{t}$ is an uncorrelated error. It is notable that the RWD provides satisfactory results in many cases \citep[see, e.g.,][]{TLB00, LM01}. From this forecast of the principal component scores, the forecast age-specific log mortality rates are obtained using the estimated age effects $a_x$ and $b_x$ in~\eqref{eq:LC}.

Note that the LC model can also be formulated within a Generalised Linear Model framework with a generalised error distribution \citep[see, e.g.,][]{Tabeau01, BDV02, RH03}. In this setting, the LC model parameters can be estimated by maximum likelihood methods based on the choice of the error distribution. Thus, in line with traditional actuarial practice, this approach assumes that the age- and period-specific number of deaths are independent realisations from a Poisson distribution. In the case of Gaussian error, the estimation method based on either the singular value decomposition or maximum likelihood method leads to the same parameter estimates.

Two sources of uncertainty are considered in the LC model: errors in the parameter estimation of the LC model and forecast errors in the forecasted principal component scores. Using a univariate time series model, we can obtain multi-step-ahead forecasts for the principal component scores, $\{\widehat{\kappa}_1,\dots,\widehat{\kappa}_n\}$. Let the $h$-step-ahead forecast errors be given by $\nu_{t,h}=\widehat{\kappa}_t-\widehat{\kappa}_{t|t-h}$ for $t=h+1,\dots,n$. These can then be sampled with replacement to give a bootstrap sample of $\kappa_{n+h}$:
\begin{equation*}
\widehat{\kappa}_{n+h|n}^{(b)} = \widehat{\kappa}_{n+h|n}+\widehat{\nu}_{\ast,h}^{(b)},\qquad b=1,\dots,B,
\end{equation*}
where $\widehat{\nu}_{\ast,h}^{(b)}$ are sampled with replacement from $\{\widehat{\nu}_{t,h}\}$.

Assuming the first principal component approximates the data $\ln \bm{m}$ relatively well, the model residual should contribute nothing but random noise. Consequently, we can bootstrap the model fit error in~\eqref{eq:LC} by sampling with replacement from the model residual term $\{\widehat{e}_{1,x},\dots,\widehat{e}_{n,x}\}$ for a given age $x$.

Adding the two components of variability, we obtain $B$ variants of $\ln m_{n+h,x}$:
\begin{equation}
\ln \widehat{m}_{n+h|n,x}^{(b)} = \widehat{\kappa}_{n+h|n}^{(b)}\widehat{b}_x + \widehat{e}_{n+h,x}^{(b)}, \label{eq:LC_PI}
\end{equation}
where $\widehat{\kappa}_{n+h|n}^{(b)}$ denotes the forecast of the bootstrapped principal component scores.

Since the LC model forecasts age-specific central mortality rate, we convert forecast mortality rate to the probability of dying via a simple approximation. The formula is given as
\begin{equation*}
\widehat{q}_{n+h|n,x} = 1 - \exp^{-\widehat{m}_{n+h|n,x}},
\end{equation*}
where $\widehat{q}_{n+h|n,x}$ denotes the probability of dying in age $x$ and year $n+h$, and $\widehat{m}_{n+h|n,x}$ denotes the forecast of central mortality rate in age $x$ and year $n+h$. With an initial life-table death count of 100,000, the forecast of $\widehat{d}_{n+h|n,x}$ can be obtained from $\widehat{q}_{n+h|n,x}$.

\subsubsection{Hyndman-Ullah (HU) method}

The HU method differs from the LC method in the following three aspects:
\begin{enumerate}
\item Instead of modelling the original mortality rates, the HU method uses a $P$-spline with a monotonic constraint to smooth log mortality rates.
\item Instead of using only one principal component and its associated scores, the HU method uses six sets of principal components and their scores.
\item Instead of forecasting each set of principal component scores by an RWD method, the HU method uses an automatic algorithm of \cite{HK08} to select the optimal model and estimate the parameters in a univariate time series forecasting method.
\end{enumerate}

\subsubsection{Random walk with and without drift}

Given the findings of linear life expectancy \citep{White02, OV02} and the debate about its continuation \citep{Bengtsson03, Lee03}, it is pertinent to compare the forecast accuracy of the CoDa method with a linear extrapolation method \citep[see][pp. 274-276 for an introduction]{AS05}. Using the centred log-ratio transformation, the linear extrapolation of $z_{t, x}$ in~\eqref{eq:ncomp} is achieved by applying the RW and RWD models for each age:
\begin{equation}
z_{t+1,x} =\zeta + z_{t,x} + e_{t+1,x}, \label{eq:RW_1}
\end{equation}
where $z_{t,x}$ represents the centred log ratio transformed data at age $x=0, 1, \dots, p$ in year $t=1,2,\dots,n-1$, $\zeta$ denotes the drift term, and $e_{t+1,x}$ denotes model error term. The $h$-step-ahead point and interval forecasts are given by
\begin{align}
\widehat{z}_{n+h|n,x} &= E\left[z_{n+h,x}|z_{1,x},\dots,z_{n,x}\right] = \zeta h + z_{n,x} \notag\\
\text{var}(\widehat{z}_{n+h|n,x}) &= \text{var}[z_{n+h,x}|z_{1,x},\dots,z_{n,x}] = \text{var}(z_{n,x}) + \text{var}(e_{n+h,x}). \label{eq:RW_2}
\end{align}
Based on~\eqref{eq:RW_1} and~\eqref{eq:RW_2}, the RW can be obtained by omitting the drift term. Computationally, the forecasts of the RW and RWD are obtained by the \verb rwf \ function in the \textit{forecast} package \citep{Hyndman17} in R \citep{Team18}. Then, by way of the back-transformation of the centred log ratio, na\"{i}ve forecasts of age-specific life-table death counts can be obtained.

\section{CoDa model fitting}\label{sec:model_fit}

For the life-table death counts, we examine the goodness-of-fit of the CoDa model to the observed data. The number of retained components $L$ in~\eqref{eq:ncomp} and~\eqref{eq:2} is determined by explaining at least 85\% of the total variation. For the Australian female and male data, the chosen number of components $L=1$ and $L=2$, respectively. We attempt to interpret the first component for the Australian female and male age-specific life-table death counts.

\begin{figure}[!htbp]
\centering
\includegraphics[width=7.3cm]{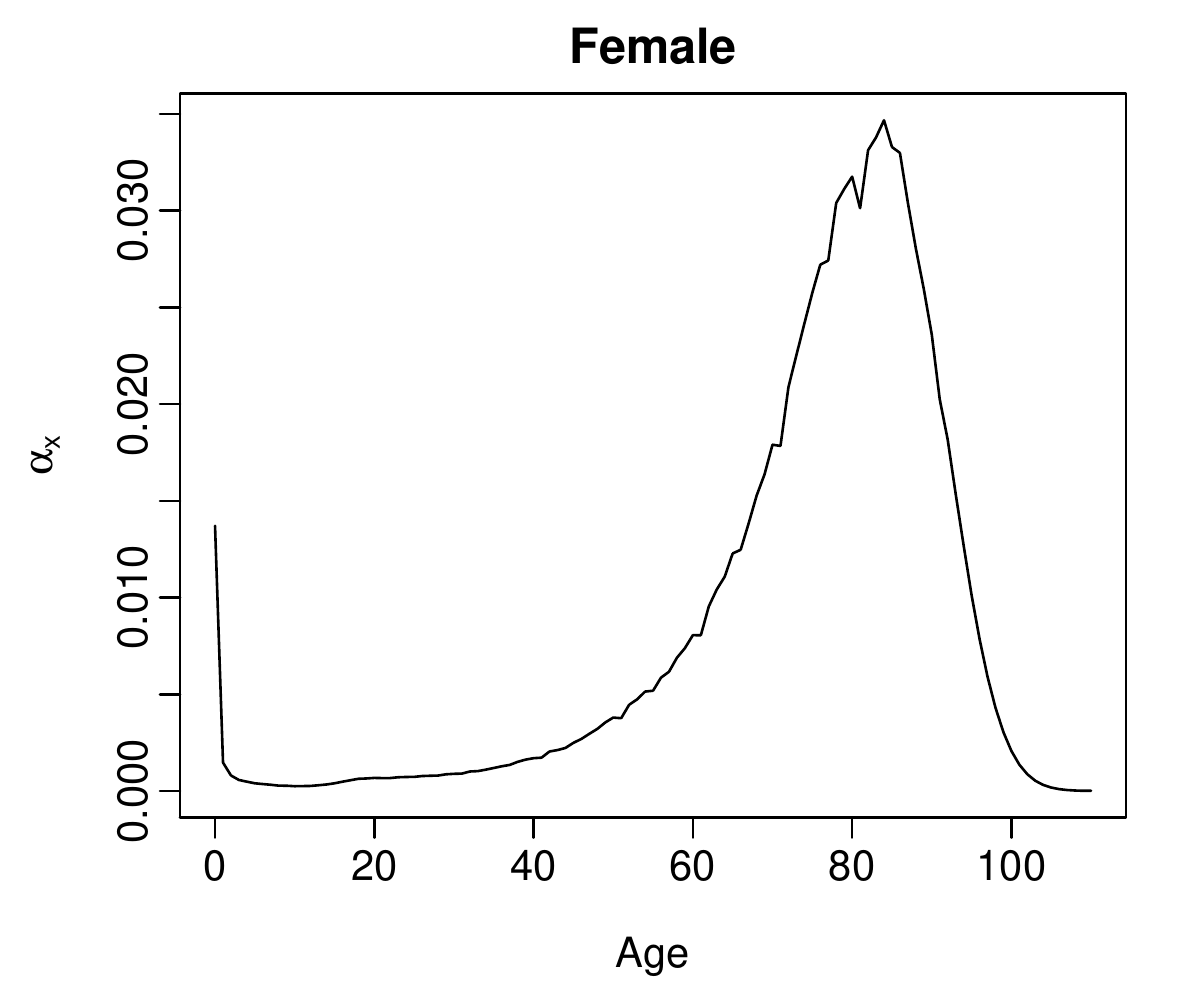}
\qquad
\includegraphics[width=7.3cm]{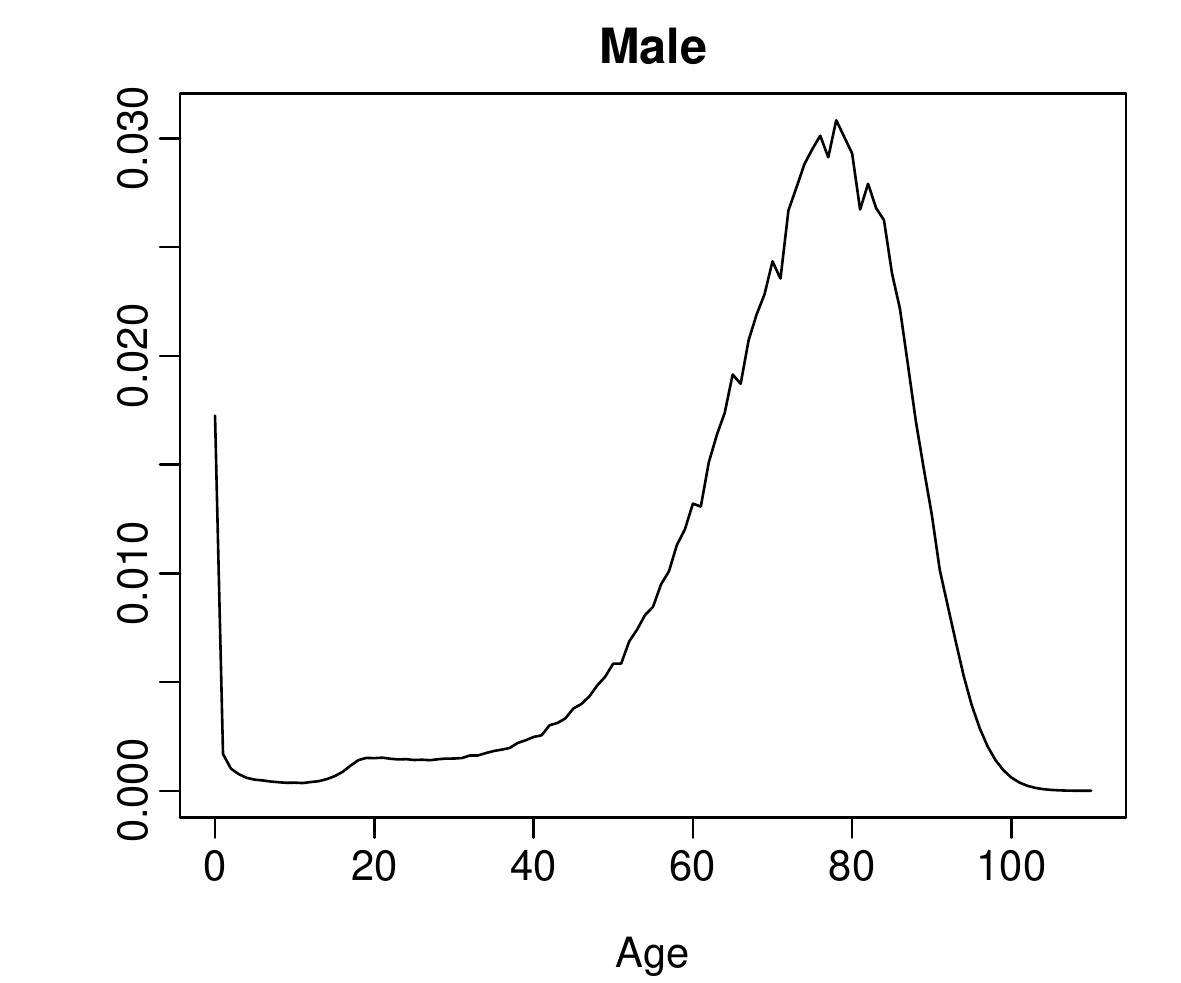}
\\
\includegraphics[width=7.3cm]{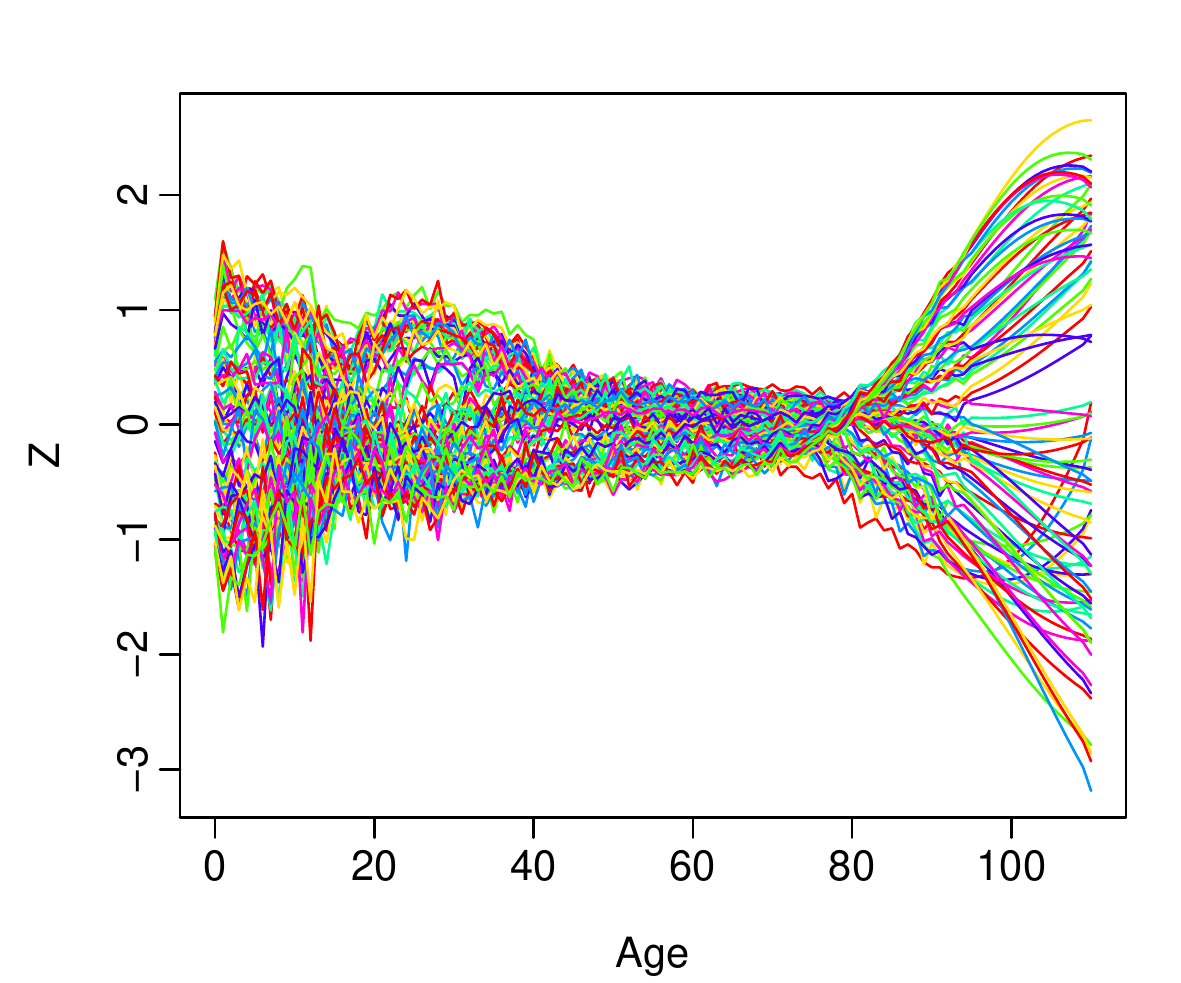}
\qquad
\includegraphics[width=7.3cm]{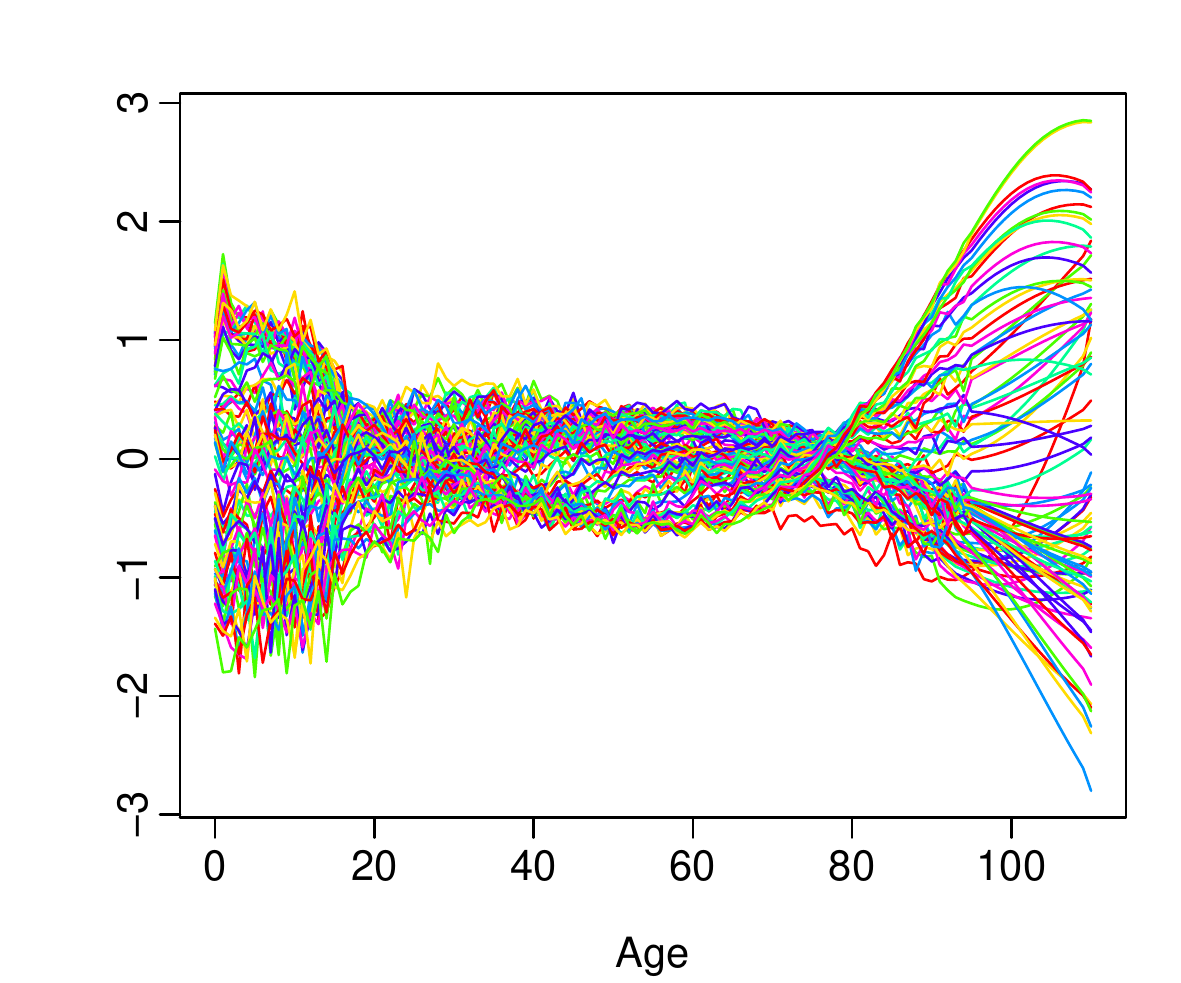}
 \\
 \includegraphics[width=7.3cm]{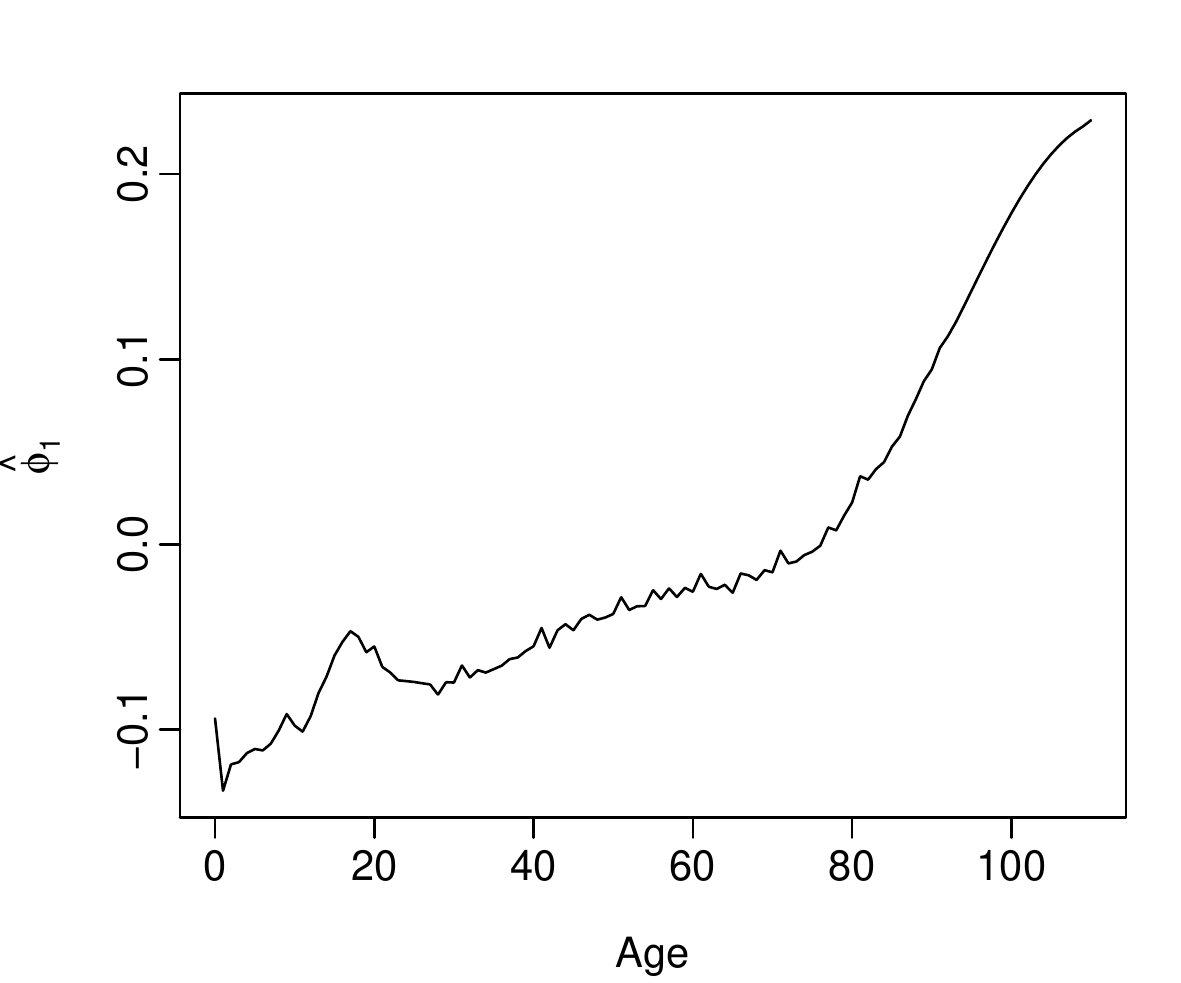}
\qquad
 \includegraphics[width=7.3cm]{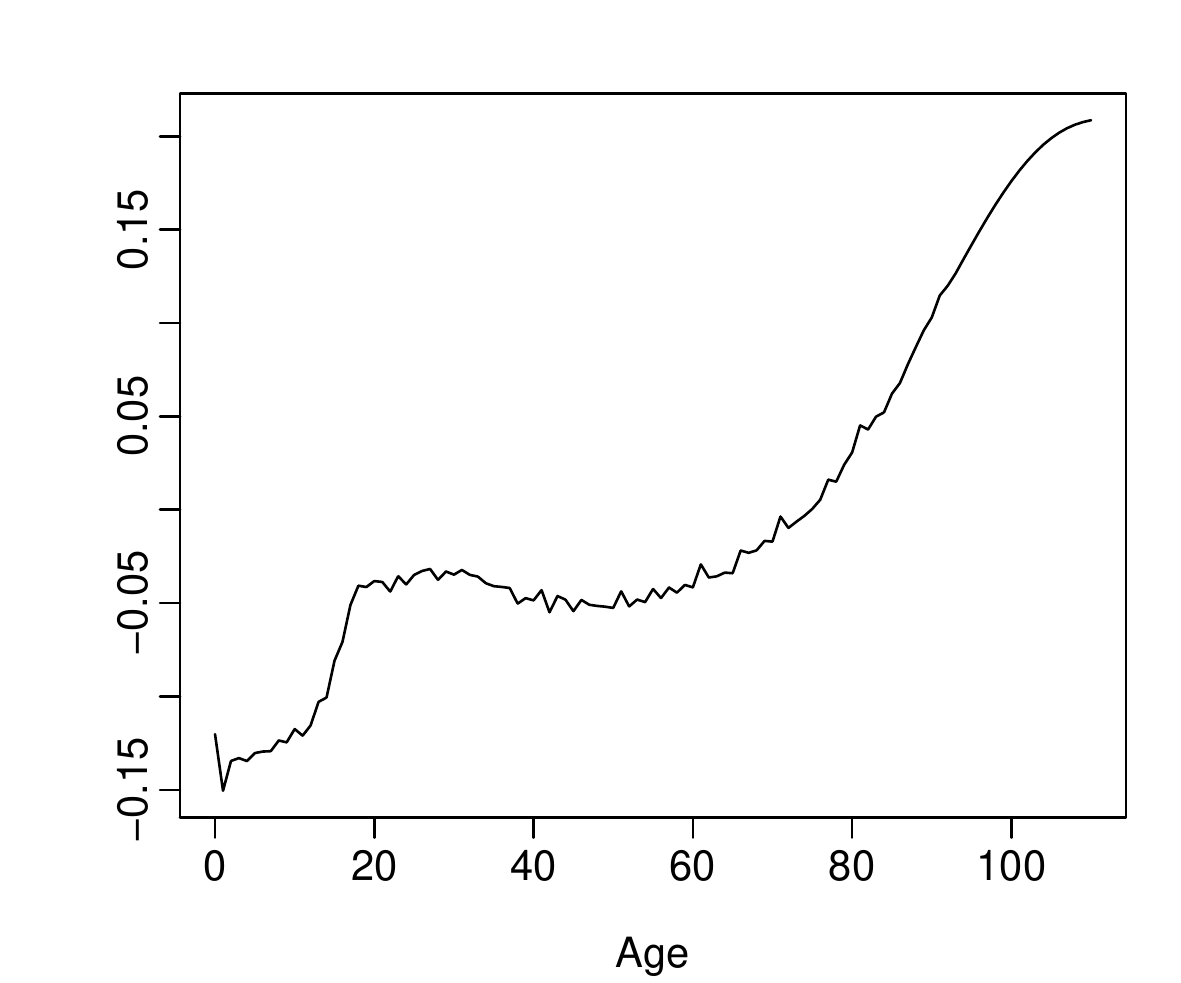}
\\
 \includegraphics[width=7.3cm]{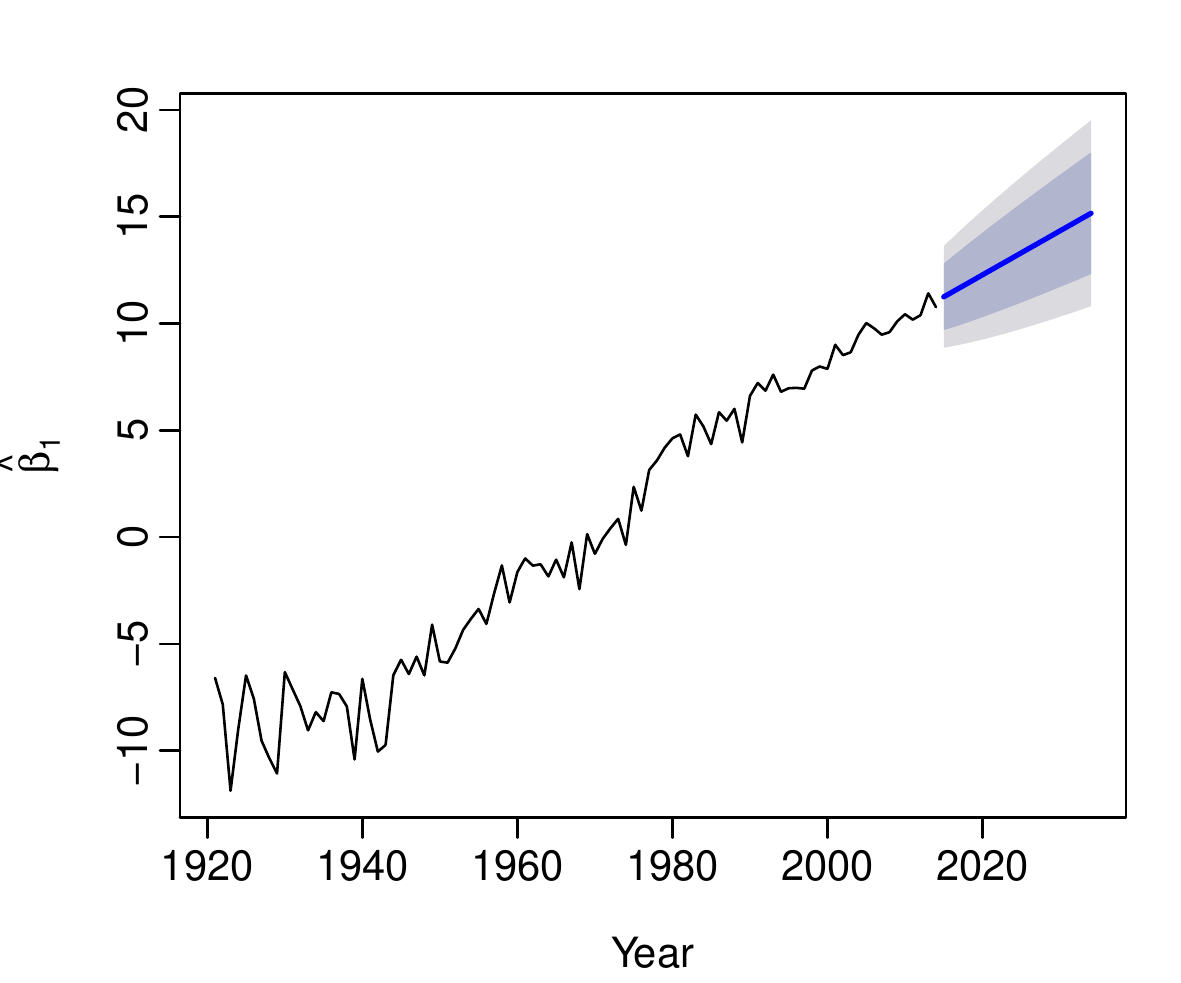}
 \qquad
 \includegraphics[width=7.3cm]{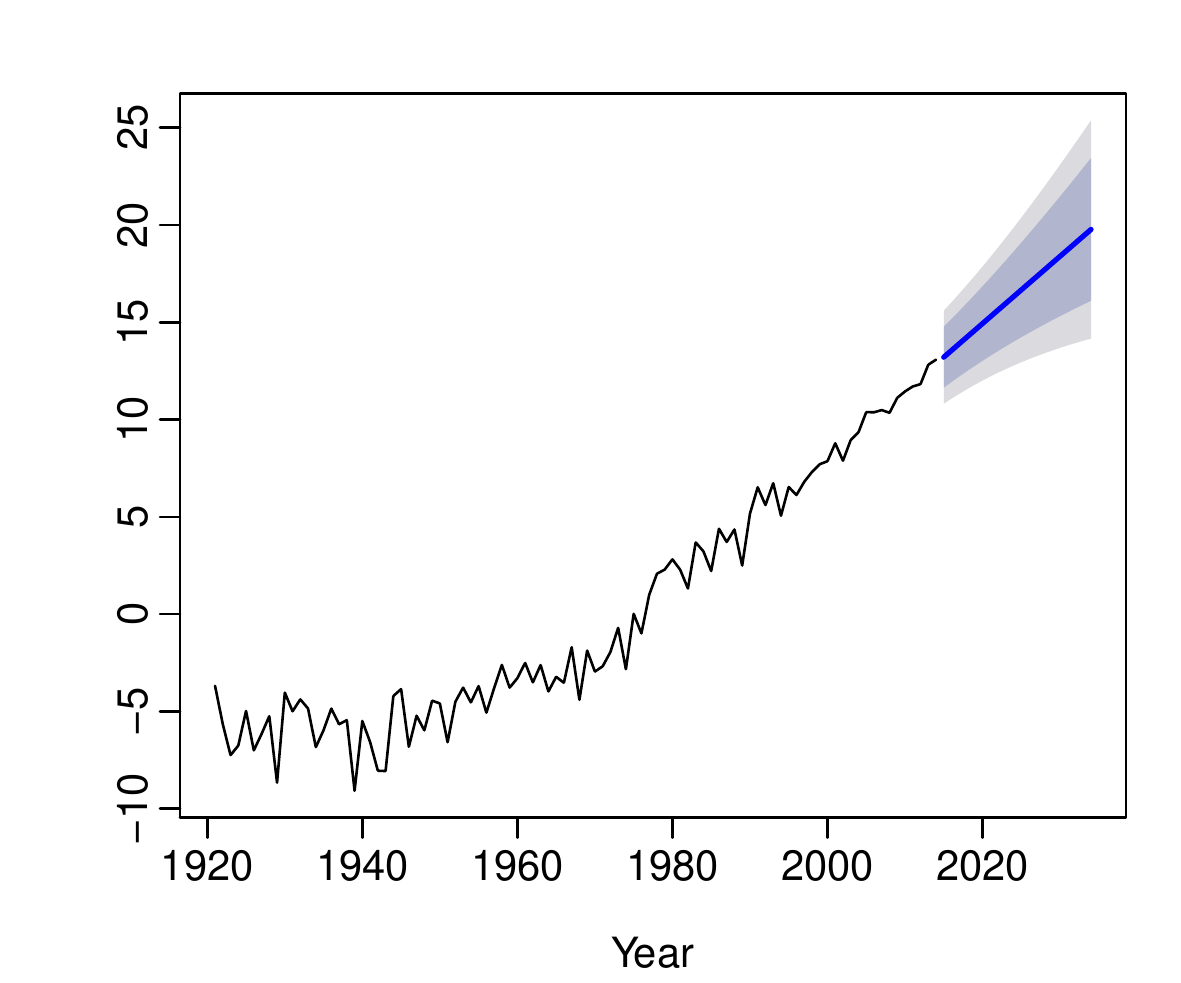}
\caption{Elements of the CoDa method for analysing the female and male age-specific life-table death counts in Australia. We present the first principal component and its scores, although we use $L=6$ for fitting.}\label{fig:model_fitting}
\end{figure}

Based on the historical death counts from 1921 to 2014 (i.e., 94 observations), in Figure~\ref{fig:model_fitting}, we present the geometric mean of female and male life-table death counts, given by $\alpha_x$, transformed data matrix $\bm{Z}$, and the first set of estimated principal component obtained by applying the principal component analysis to $\bm{Z}$. The shape of the first estimated principal component appears to be similar to the first estimated principal component in the \citeauthor{LC92}'s \citeyearpar{LC92} and \citeauthor{HU07}'s \citeyearpar{HU07} methods. Because $\bm{Z}$ is unbounded, the principal component analysis captures a similar projection direction. By using the automatic ETS forecasting method, we produce the 20-steps-ahead forecast of principal component scores for the year between 2015 and 2034.

Apart from the graphical display, we measure the in-sample goodness-of-fit via an $R^2$ criterion given as
\begin{equation*}
R^2 = 1- \frac{\sum^{111}_{x=1}\sum^{94}_{t=1}\left(d_{t,x} - \widehat{d}_{t,x}\right)^2}{\sum^{111}_{x=1}\sum^{94}_{t=1}\left(d_{t,x} - \overline{d}_x\right)^2}, 
\end{equation*}
where $d_{t,x}$ denotes the observed age-specific life-table death count for age $x$ in year $t$, and $\widehat{d}_{t,x}$ denotes the fitted age-specific life-table death count obtained from the CoDa model. For the Australian female and male data, the $R^2$ values are 0.9946 and 0.9899 using the CPV criterion while the $R^2$ values are 0.9987 and 0.9987 based on $L=6$, respectively. Using the CPV criterion, we select $L=1$ for both female and male data. One remark is that choosing the number of principal components $L=6$ does not improve greatly the $R^2$ values, but it improves the forecast accuracy greatly as will be shown in Table~\ref{tab:CoDa_results}.

\section{Comparisons of point and interval forecast accuracy}\label{sec:4}

\subsection{Forecast error criteria}\label{sec:forecast_error}

An expanding window analysis of a time-series model is commonly used to assess model and parameter stability over time, and prediction accuracy. The expanding window analysis determines the constancy of a model's parameter by computing parameter estimates and their resultant forecasts over an expanding window of a fixed size through the sample \citep[For details,][pp. 313-314]{ZW06}. Using the first 74 observations from 1921 to 1994 in the Australian female and male age-specific life-table death counts, we produce one- to 20-step-ahead forecasts. Through a rolling-window approach, we re-estimate the parameters in the time series forecasting models using the first 75 observations from 1921 to 1995. Forecasts from the estimated models are then produced for one- to 19-step-ahead. We iterate this process by increasing the sample size by one year until reaching the end of the data period in 2014. This process produces 20 one-step-ahead forecasts, 19 two-step-ahead forecasts, $\dots$, and one 20-step-ahead forecast. We compare these forecasts with the holdout samples to determine the out-of-sample point forecast accuracy. 

To evaluate the point forecast accuracy, we consider the MAPE which measures how close the forecasts are in comparison to the actual values of the variable being forecast, regardless of the direction of forecast errors. The error measure can be written as
\begin{equation*}
\text{MAPE}(h) = \frac{1}{111\times (21-h)}\sum^{20}_{\varsigma=h}\sum^{111}_{x=1}\left|\frac{d_{n+\varsigma, x}-\widehat{d}_{n+\varsigma,x}}{d_{n+\varsigma, x}}\right|\times 100,
\end{equation*}
where $d_{n+\varsigma,x}$ denotes the actual holdout sample for the $x^{\text{th}}$ age and $\varsigma^{\text{th}}$ forecasting year, while $\widehat{d}_{n+\varsigma,x}$ denotes the point forecasts for the holdout sample. 

To evaluate and compare the interval forecast accuracy, we consider the interval score of \cite{GR07}. For each year in the forecasting period, the $h$-step-ahead prediction intervals are calculated at the $100(1-\gamma)\%$ nominal coverage probability. We consider the common case of the symmetric $100(1-\gamma)\%$ prediction intervals, with lower and upper bounds that are predictive quantiles at $\gamma/2$ and $1-\gamma/2$, denoted by $\widehat{d}^l_{n+\varsigma, x}$ and $\widehat{d}^u_{n+\varsigma, x}$. As defined by \cite{GR07}, a scoring rule for the interval forecasts at time point $d_{n+\varsigma,x}$ is 
\begin{align*}
S_{\gamma,\varsigma}\left[\widehat{d}_{n+\varsigma, x}^l, \widehat{d}_{n+\varsigma, x}^u, d_{n+\varsigma, x}\right]=\left(\widehat{d}^u_{n+\varsigma, x} - \widehat{d}^l_{n+\varsigma, x}\right)&+\frac{2}{\gamma}\left(\widehat{d}^l_{n+\varsigma, x} - d_{n+\varsigma, x}\right)\mathds{1}\left\{d_{n+\varsigma, x} < \widehat{d}_{n+\varsigma, x}^l\right\} \\
& + \frac{2}{\gamma}\left(d_{n+\varsigma, x}-\widehat{d}_{n+\varsigma, x}^u\right)\mathds{1}\left\{d_{n+\varsigma, x}>\widehat{d}_{n+\varsigma, x}^u\right\}, 
\end{align*}
where $\mathds{1}\{\cdot\}$ represents the binary indicator function, and $\gamma$ denotes the level of significance, customarily $\gamma=0.2$ or $\gamma=0.05$. The interval score rewards a narrow prediction interval, if and only if the true observation lies within the prediction interval. The optimal interval score is achieved when $d_{n+\varsigma, x}$ lies between $\widehat{d}_{n+\varsigma,x}^l$ and $\widehat{d}_{n+\varsigma, x}^u$, and the distance between $\widehat{d}_{n+\varsigma, x}^l$ and $\widehat{d}_{n+\varsigma, x}^u$ is minimal.

For different ages and years in the forecasting period, the mean interval score is defined by
\begin{equation*}
\overline{S}_{\gamma}(h) = \frac{1}{111\times (21-h)}  \sum_{\varsigma = h}^{20} \sum_{x=1}^{111} S_{\gamma,\varsigma}\left[\widehat{d}_{n+\varsigma,x}^{l}, \widehat{d}_{n+\varsigma,x}^{u}; d_{n+\varsigma,x}\right],
\end{equation*}
where $S_{\gamma,\varsigma}\left[\widehat{d}_{n+\varsigma,x}^{l}, \widehat{d}_{n+\varsigma,x}^{u}; d_{n+\varsigma,x}\right]$ denotes the interval score at the $\varsigma^{\text{th}}$ curve in the forecasting year.

\subsection{Forecast results}

Using the expanding window approach, we compare the point and interval forecast accuracies among the CoDa, HU, LC and two na\"{i}ve RW methods based on the MAPE and mean interval score. Also, we consider forecasting each set of the estimated principal component scores by the ARIMA, ETS, RWD and RW. The overall point and interval forecast error results are presented in Table~\ref{tab:CoDa_results}, where we average over the 20 forecast horizons. From Table~\ref{tab:CoDa_results}, we find that the CoDa method with the ETS forecasting method generally performs the best among a range of methods considered. Among the four univariate time series forecasting methods, the ETS method produces the smallest errors and thus is recommended to be used in practice. Also, setting $L=6$, including additional principal component decomposition pairs, generally provides more accurate point and interval forecast accuracies than setting $L=1$ or $L=2$ for the Australian female and male data. 

\begin{center}
\tabcolsep 0.09in
\begin{longtable}{@{}llllrrrrr@{}}
\caption{A comparison of the point and interval forecast accuracy, as measured by the overall MAPE and mean interval score, among the CoDa, HU, LC and two na\"{i}ve RW methods (in italic) using the holdout sample of the Australian female and male data. Further, we consider four univariate time series forecasting methods for the CoDa and HU methods. In terms of interval forecast accuracy, we compare the finite-sample performance between the proposed bootstrap approach and an existing bootstrap approach of \cite{BC17}. The smallest errors are highlighted in bold.}\label{tab:CoDa_results}\\
\toprule
Series & $L$ & Method & Forecasting  & \multicolumn{5}{c}{Criteria} \\
& & &  method &        & \multicolumn{2}{c}{Proposed bootstrap} & \multicolumn{2}{c}{Existing bootstrap} \\
& & & & MAPE & $\overline{S}_{\gamma=0.2}$ & $\overline{S}_{\gamma=0.05}$ & $\overline{S}_{\gamma=0.2}$ & $\overline{S}_{\gamma=0.05}$ \\\midrule
\endfirsthead
\toprule
Series & $L$ & Method & Forecasting & \multicolumn{5}{c}{Criteria} \\
& & &  method &        & \multicolumn{2}{c}{Proposed bootstrap} & \multicolumn{2}{c}{Existing bootstrap} \\
& & & & MAPE & $\overline{S}_{\gamma=0.2}$ & $\overline{S}_{\gamma=0.05}$ & $\overline{S}_{\gamma=0.2}$ & $\overline{S}_{\gamma=0.05}$ \\\midrule
\endhead
\hline \multicolumn{9}{r}{{Continued on next page}} \\
\endfoot
\endlastfoot
Female & CPV & CoDa & ARIMA & 20.51 & 423.69 & 653.17 & 575.89 & 670.54 \\
	     & 	         &          & ETS     & 20.65 & 492.77 & 870.10 & 529.32 & 655.10 \\
	     &		&	     & RWD   & 21.32 & 476.77 & 782.52 & 711.58 & 859.78 \\
	     & 		& 	     & RW      & 32.18 & 512.30 & 629.45 & 1302.91 & 2368.32 \\
	     \\
	     & $L=6$ & CoDa & ARIMA & 17.02 & 298.93 & 431.48 & 578.12 & 812.75 \\
	     &		&		& ETS & \textBF{14.60} & \textBF{232.10} & \textBF{369.76} & 277.51 & 407.85 \\
	     & 		& 		& RWD & 16.46 & 369.77 & 640.87 & 427.40 & 689.46	\\
	     & 		& 		& RW & 29.25 & 985.94 & 2512.42 & 1056.94 & 2422.63		\\
	     \\
	     	     & 	        & HU        & ARIMA  & 19.60 & 348.84 & 462.21 \\
	     &		& 		& ETS      & 16.71 & 262.36 & 376.59 \\
	     &		&		& RWD 	& 16.49 & 349.00 & 472.72 \\
	     &		& 		& RW 	& 30.95 & 350.32 & 442.68 \\
\cmidrule{2-9}
	    & 		& LC 	& 		& 26.54 & 516.79 & 667.54 \\	
	    &		& \textit{RWD}	& 		& 16.59 & 703.09 & 1185.52 \\
	    &		& \textit{RW}	& 		& 30.04 & 703.79 & 1186.35 \\
	    \midrule
Male     & CPV  & CoDa  & ARIMA & 31.93  & 1122.74 & 2079.19 & 1624.23 & 3889.32 \\
	&		&		& ETS &  25.48 &  644.38 & 752.16 & 1139.60 & 2184.71 \\
	&		& 		& RWD & 32.16  & 690.67 & 896.75 & 1750.66 & 4434.10 \\
	&		& 		& RW & 44.82 &  1428.70 & 3142.54 & 2115.80 & 5683.19 \\	      
\\
	& $L=6$ & CoDa & ARIMA & 27.18 & 648.85 & 860.78 & 1484.00 & 4067.55 \\
	&		&	& ETS & \textBF{18.37} & \textBF{371.22} & \textBF{516.23} & 699.92 & 1417.62 \\
	&		&	& RWD & 23.59  & 776.14 & 1221.44 & 1124.12 & 2854.69 \\
	&		&	& RW & 37.48 & 457.71 & 576.84 & 1638.70 & 4721.93 \\ 	
\\
	     & 	        & HU        & ARIMA  & 27.22 & 641.30 & 944.56 \\
	     &		& 		& ETS      & 24.64 &  726.86 & 1281.29 \\
	     &		&		& RWD 	& 25.14 & 806.16 & 1703.06 \\
	     &		& 		& RW 	& 39.48 &  965.43 & 2288.80 \\
\cmidrule{2-9}
	     &		& LC 		&	& 38.61 &  1273.36 & 2692.18 \\
	     &		& \textit{RWD}	& 		& 23.60 & 783.15 & 1166.38 \\
	    &		& \textit{RW}	& 		& 38.18 & 783.30 & 1166.34 \\     
\bottomrule
\end{longtable}
\end{center}

The CoDa method tends to provide less bias forecasts than the Lee-Carter method as it allows the rate of mortality improvements to change over time \citep[see][]{BC17}. With respect to the positivity and summability constraints, the CoDa method can adopt temporal changes of the age distribution of the life-table death counts over the years. The unsatisfactory performance of the HU and LC methods may because of the approximation of converting the forecasts of central mortality rate to the probability of dying at higher ages. The inferior performance of the two na\"{i}ve RW methods may because of their slow responses to adapt to the change of age distribution of death counts. Between the two na\"{i}ve RW methods, we found that the RW is preferable for producing point forecasts, while the RWD is preferable for producing interval forecasts.

\begin{figure}[!htbp]
\centering
\includegraphics[width=8.25cm]{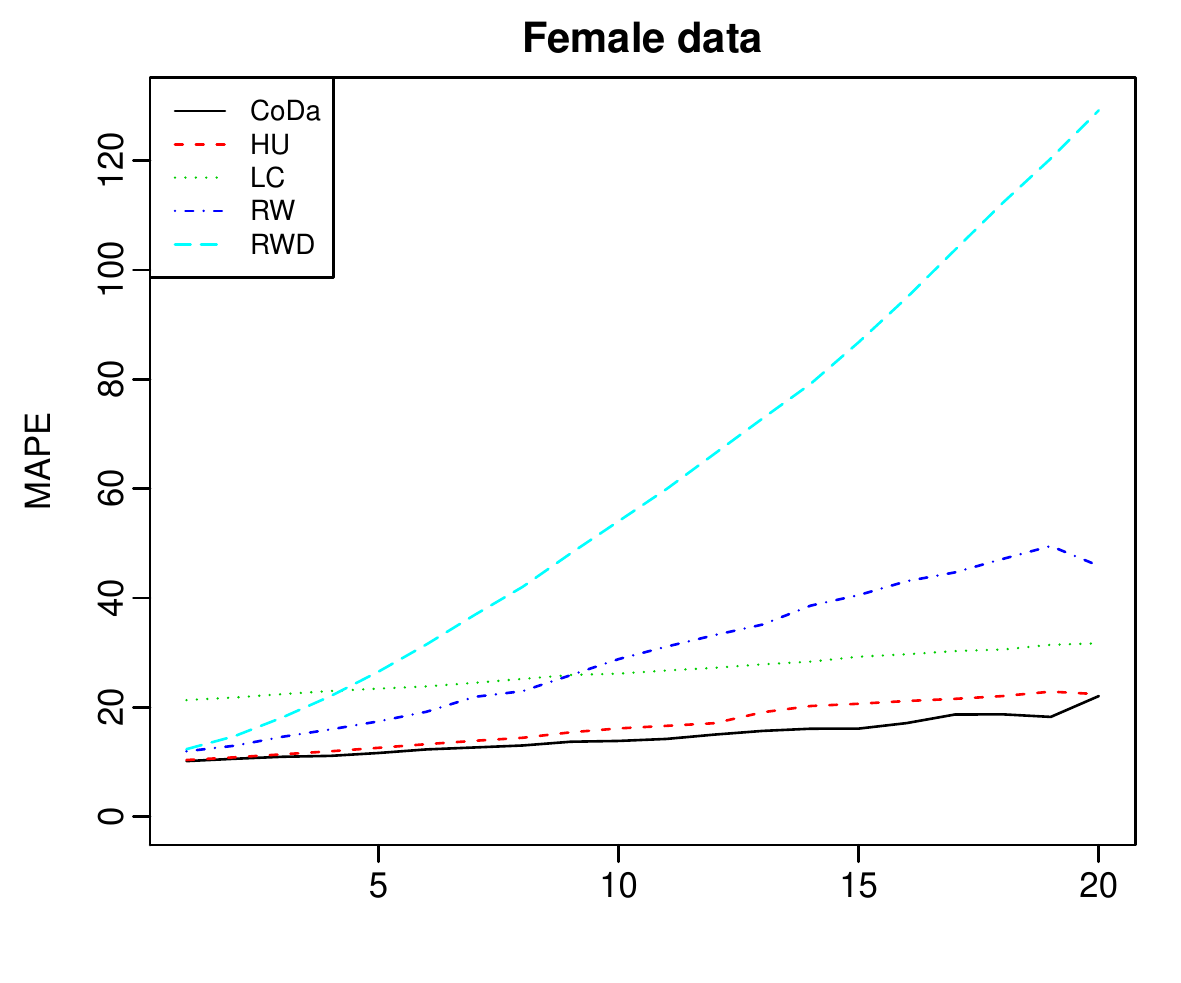}
\qquad
\includegraphics[width=8.25cm]{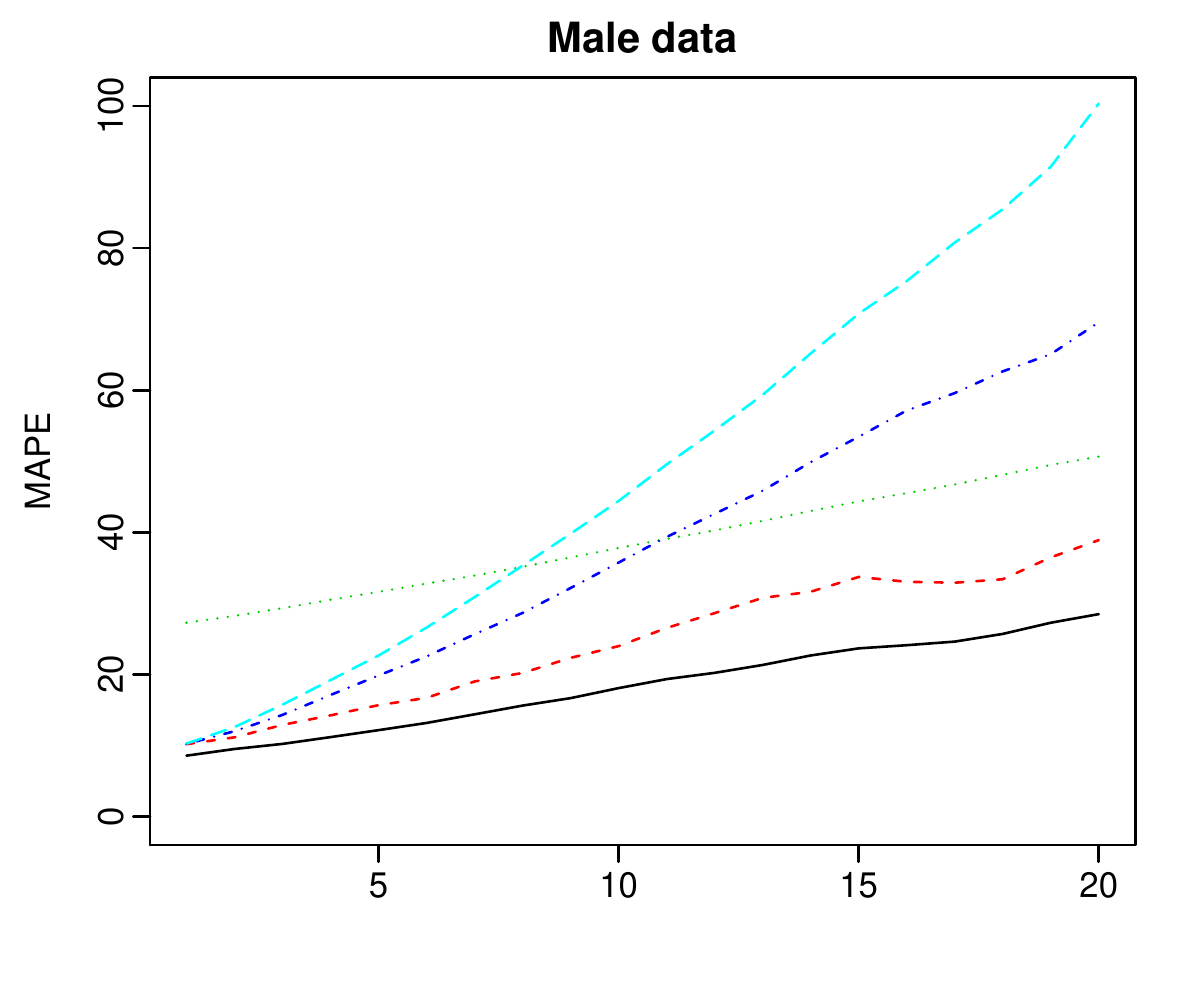}
\\
\includegraphics[width=8.25cm]{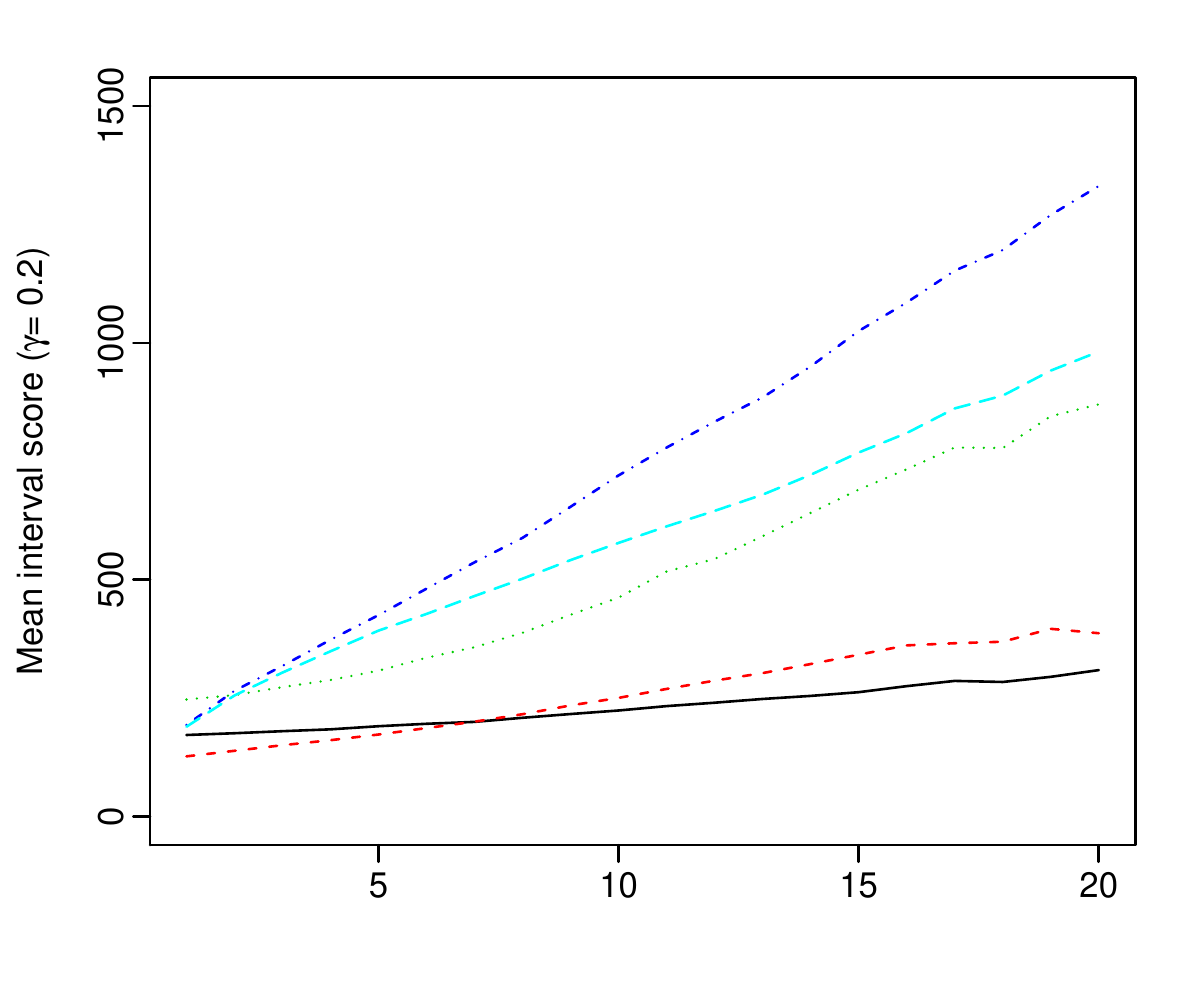}
\qquad
\includegraphics[width=8.25cm]{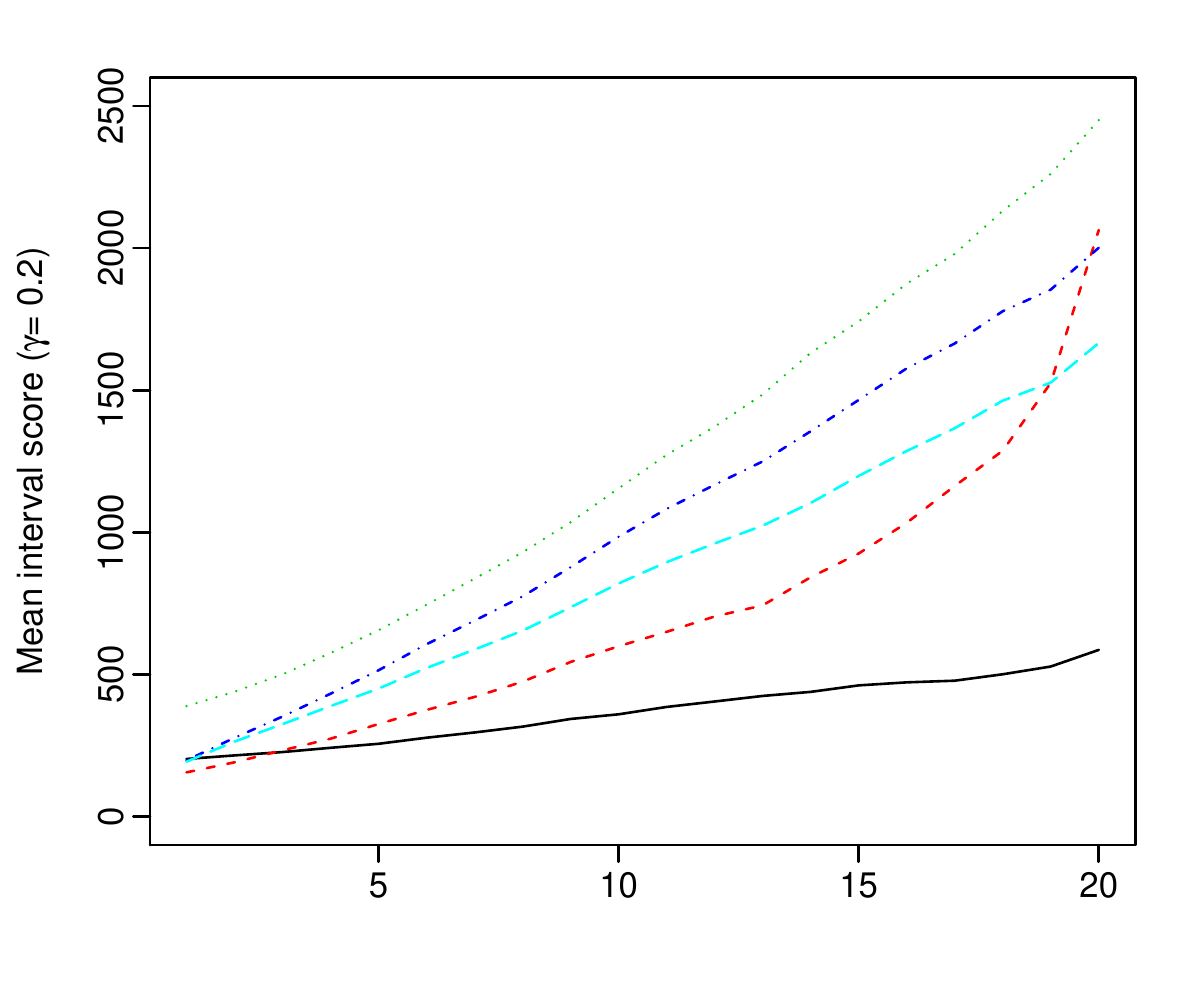}
\\
\includegraphics[width=8.25cm]{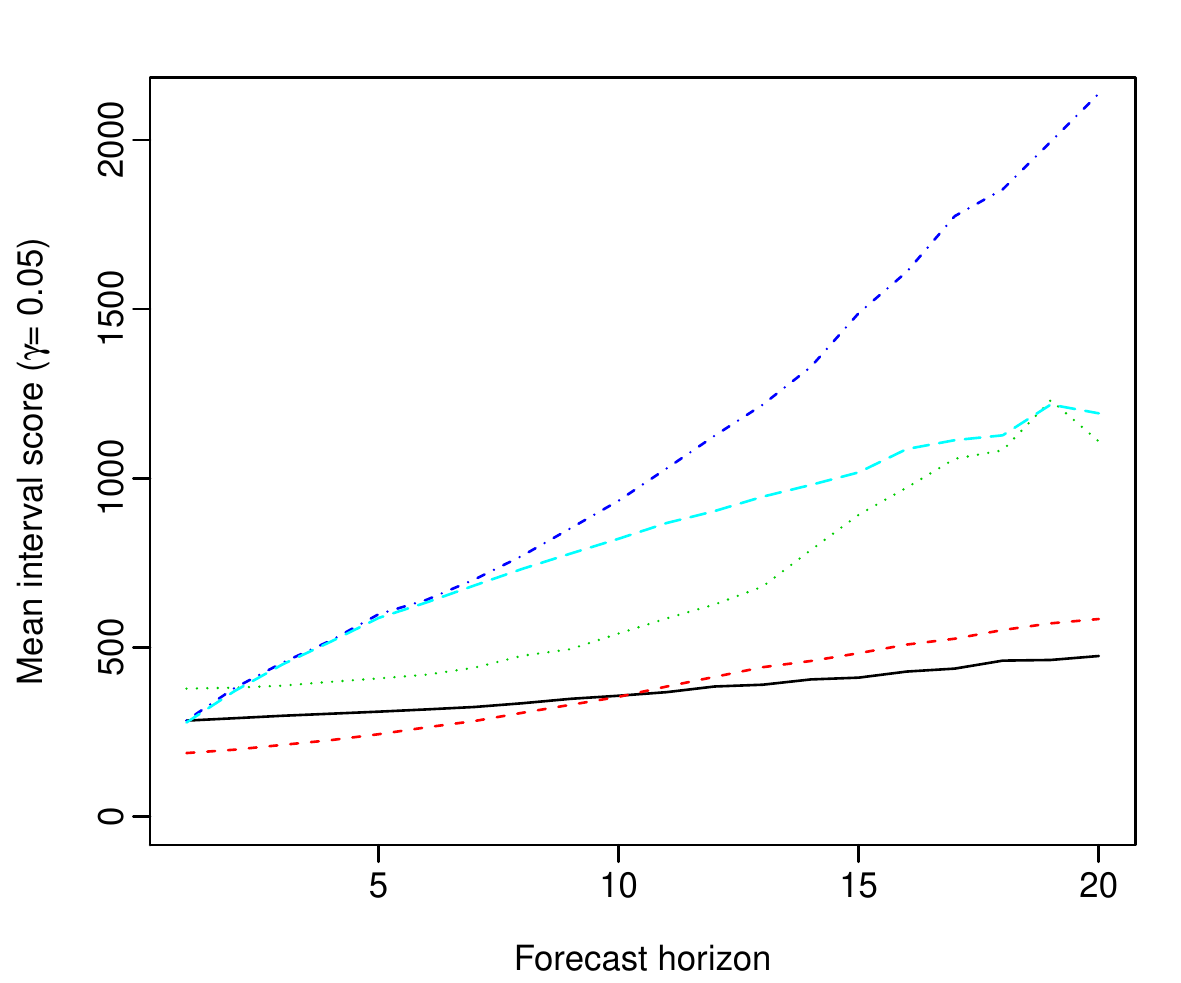}
\qquad
\includegraphics[width=8.25cm]{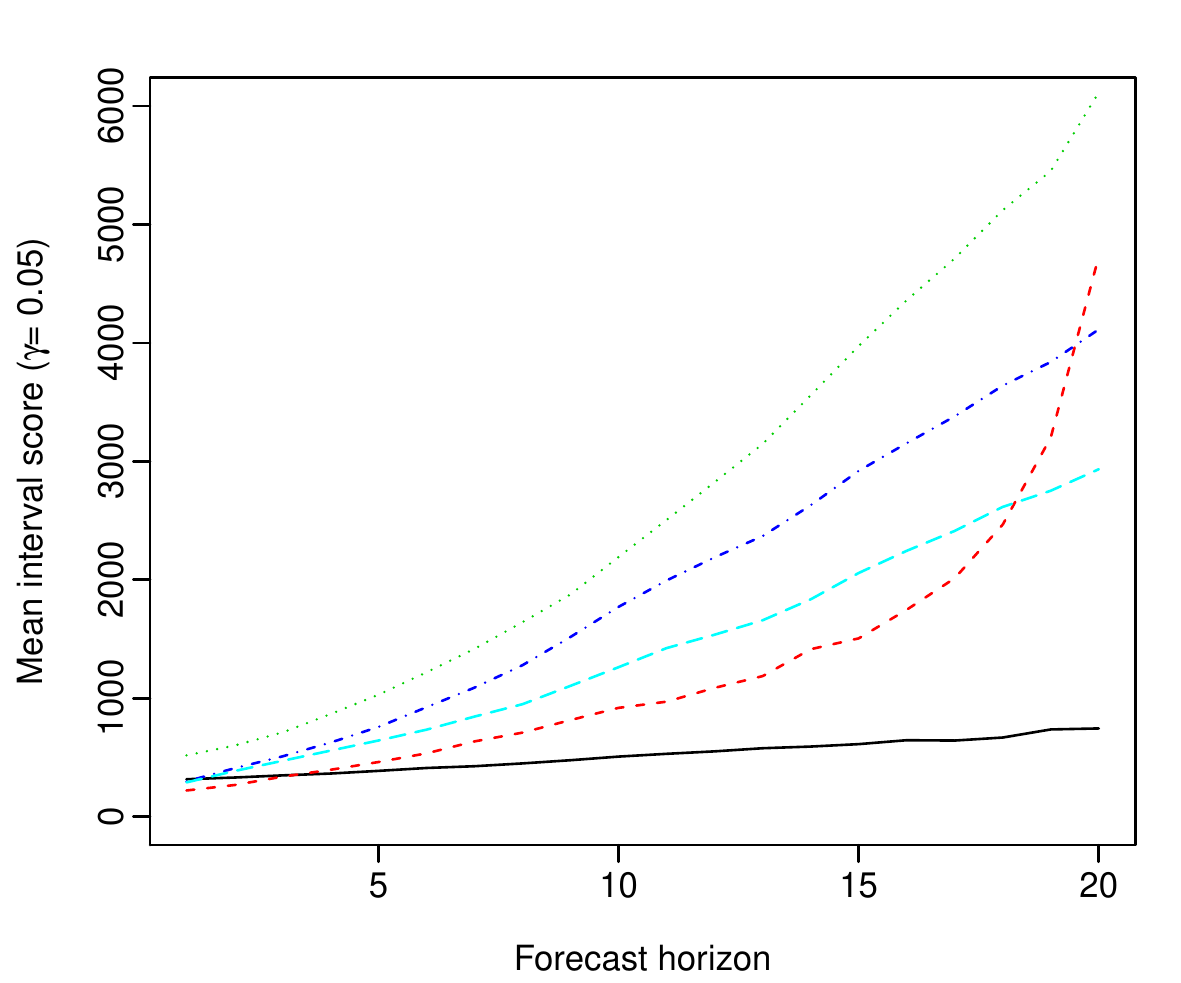}
\caption{A comparison of the point and interval forecast accuracy, as measured by the MAPE and mean interval score, among the CoDa, HU, LC and two na\"{i}ve RW methods using the holdout sample of the Australian female and male data. In the CoDa method, we use the ETS forecasting method with the number of principal components $L=6$.}\label{fig:point_accuracy}
\end{figure}

While Table~\ref{tab:CoDa_results} presents the average over 20 forecast horizons, we show the one-step-ahead to 20-step-ahead point and interval forecast errors in Figure~\ref{fig:point_accuracy}. Since it is advantageous to set $L=6$, we report the CoDa method with the ETS forecasting method and $L=6$ in Figure~\ref{fig:point_accuracy}. The difference in forecast accuracy between the CoDa and the other methods is widening over the forecast horizon. We suspect that in the relatively longer forecast horizon, the errors associated with all the methods become larger, but they are relatively smaller for the CoDa method with the ETS forecasting method and $L=6$.

\section{\mbox{Application to a single-premium temporary immediate annuity}}\label{sec:6}

An important use of mortality forecasts for those individuals at ages over 60 is in the pension and insurance industries, whose profitability and solvency crucially rely on accurate mortality forecasts to appropriately hedge longevity risks. When a person retires, an optimal way of guaranteeing one individual's financial income in retirement is to purchase an annuity \citep[as demonstrated by][]{Yaari65}. An annuity is a financial contract offered by insurers guaranteeing a steady stream of payments for either a temporary or the lifetime of the annuitants in exchange for an initial premium fee.

Following \cite{SH17b}, we consider temporary annuities, which have grown in popularity in a number of countries (e.g., Australia and United States of America), because lifetime immediate annuities, where rates are locked in for life, have been shown to deliver poor value for money \citep[i.e., they may be expensive for the purchasers; see for example][Chapter 6]{CT08}. These temporary annuities pay a pre-determined and guaranteed level of income which is higher than the level of income provided by a lifetime annuity for a similar premium. Fixed-term annuities offer an alternative to lifetime annuities and allow the purchaser the option of also buying a deferred annuity at a later date. 

\begin{figure}[!htbp]
\centering
\includegraphics[width=12cm]{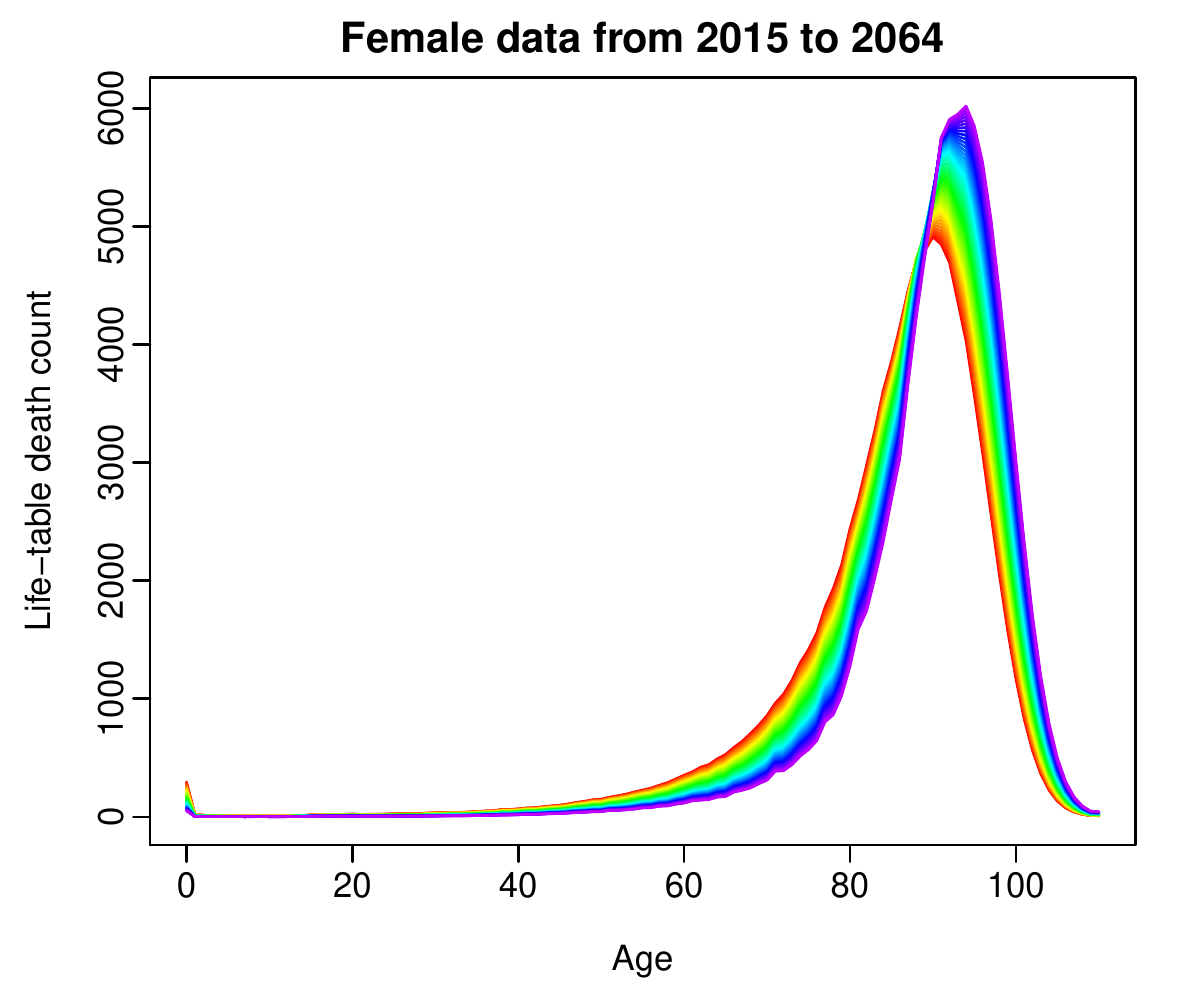}
\caption{Age-specific life-table death count forecasts from 2015 to 2064 for Australian females.}\label{fig:mortality_fore}
\end{figure}

Using the CoDa method, we obtain forecasts of age-specific life-table death counts and then determine the corresponding survival probabilities. In Figure~\ref{fig:mortality_fore}, we present the age-specific life-table death count forecasts from 2015 to 2064 for Australian females.

With the mortality forecasts, we then input the forecasts of death counts to the calculation of single-premium term immediate annuities \citep[see][p. 114]{DHW09}, and we adopt a cohort approach to the calculation of the survival probabilities. The $\tau$ year survival probability of a person aged $x$ currently at $t=0$ (or year 2016) is determined by
\begin{align*}
_{\tau}p_x = \prod^{\tau}_{j=1} p_{x+j-1} 
=\prod^{\tau}_{j=1} \left(1 - q_{x+j-1}\right) = \prod^{\tau}_{j=1} \left(1 - \frac{d_{x+j-1}}{l_{x+j-1}}\right),
\end{align*}
where $d_{x+j-1}$ denotes the number of death counts between ages $x+j-1$ and $x+j$; and $l_{x+j-1}$ denotes the number of lives alive at age $x+j-1$. Note that ${}_{\tau}p_x$ is a random variable given that death counts for $j=1,\dots,\tau$ are the forecasts obtained by the CoDa method.

The price of an annuity with a maturity term of a $T$ year is a random variable, as it depends on the value of zero-coupon bond price and future mortality. The annuity price can be written for an $x$-year-old with benefit one Australian dollar per year is given by
\begin{equation*}
a_x^T = \sum^T_{\tau=1}B(0,\tau){}_{\tau}p_x,
\end{equation*}
where $B(0,\tau)$ is the $\tau$-year bond price and ${}_{\tau}p_x$ denotes the survival probability. 

In Table~\ref{tab:annuity_tab}, to provide an example of the annuity calculations, we compute the best estimate of the annuity prices for different ages and maturities for a female policyholder residing in Australia. We assume a constant interest rate at $\eta=3\%$ and hence zero-coupon bond is given as $B(0, \tau) = \exp^{-\eta \tau}$. Although the difference in annuity price might appear to be small, any mispricing can involve a significant risk when considering a large annuity portfolio. Given that an annuity portfolio consists of $N$ policies where the benefit per year is $B$, any underpricing of $\tau\%$ of the actual annuity price will result in a shortfall of $NBa_x^T\tau/100$, where $a_x^T$ is the estimated annuity price being charged with benefit one Australian dollar per year. For example $\tau=0.1\%$, $N=10,000$ policies written to 80-year-old female policyholders with maturity $\tau=20$ years and benefit $20,000$ Australian dollars per year will result in a shortfall of $10,000\times 20,000\times 7.8109  \times 0.1\% =  1.5622$ million.

\begin{table}[!htbp]
\centering
\tabcolsep 0.27in
  \caption{Estimates of annuity prices with different ages and maturities ($T$) for a female policyholder residing in Australia. These estimates are based on forecast mortality rates from 2015 to 2064. We consider only contracts with maturity so that age + maturity $\leq 110$. If age + maturity $> 110$, NA will be shown in the table.}\label{tab:annuity_tab} 
{\renewcommand{\arraystretch}{0.98}
\begin{tabular}{@{}lrrrrrr@{}}
\toprule
 Age & $T=5$ & $T=10$ & $T=15$ & $T=20$ & $T=25$ & $T=30$ \\ 
\midrule
\underline{CPV} & \\
60 & 4.5109 & 8.2748 & 11.3625 & 13.8221 & 15.6564 & 16.8572 \\ 
  65 & 4.4831 & 8.1608 & 11.0904 & 13.2752 & 14.7054 & 15.4242 \\ 
  70 & 4.4314 & 7.9614 & 10.5940 & 12.3173 & 13.1833 & 13.4609 \\ 
  75 & 4.3445 & 7.5847 & 9.7056 & 10.7715 & 11.1132 & 11.1684 \\ 
  80 & 4.1440 & 6.8566 & 8.2198 & 8.6568 & 8.7274 & 8.7322 \\ 
  85 & 3.7975 & 5.7060 & 6.3177 & 6.4166 & 6.4232 & NA \\ 
  90 & 3.1874 & 4.2090 & 4.3742 & 4.3852 & NA & NA \\ 
  95 & 2.4261 & 2.8183 & 2.8445 & NA & NA & NA \\ 
  100 & 1.6751 & 1.7870 & NA & NA & NA & NA \\ 
  105 & 1.1283 & NA & NA & NA & NA & NA \\
\\
  \underline{$L=6$} & \\
  60 & 4.5197 & 8.3091 & 11.4419 & 13.9673 & 15.8921 & 17.1909 \\ 
  65 & 4.4972 & 8.2151 & 11.2122 & 13.4965 & 15.0380 & 15.8363 \\ 
  70 & 4.4538 & 8.0441 & 10.7806 & 12.6271 & 13.5835 & 13.8961 \\ 
  75 & 4.3788 & 7.7163 & 9.9683 & 11.1348 & 11.5160 & 11.5764 \\ 
  80 & 4.2042 & 7.0411 & 8.5105 & 8.9907 & 9.0668 & 9.0715 \\ 
  85 & 3.8680 & 5.8714 & 6.5261 & 6.6299 & 6.6364 & NA \\ 
  90 & 3.2348 & 4.2919 & 4.4595 & 4.4699 & NA & NA \\ 
  95 & 2.4406 & 2.8276 & 2.8515 & NA & NA & NA \\ 
  100 & 1.6466 & 1.7486 & NA & NA & NA & NA \\ 
  105 & 1.0859 & NA & NA & NA & NA & NA \\ 
 \bottomrule
\end{tabular}}
\end{table}

To measure forecast uncertainty, we construct the bootstrapped age-specific life-table death counts, derive the survival probabilities and calculate the corresponding annuities associated with different ages and maturities. Given that we consider ages from 60 to 110, we construct 50-steps-ahead bootstrap forecasts of age-specific life-table death counts. In Table~\ref{tab:annuity_tab_int}, we present the 95\% pointwise prediction intervals of annuities for different ages and maturities, where age + maturity $\leq 110$.

\begin{table}[!htbp]
\centering
\tabcolsep 0.015in
  \caption{95\% pointwise prediction intervals of annuity prices with different ages and maturities ($T$) for a female policyholder residing in Australia. These estimates are based on forecast mortality rates from 2015 to 2064. We consider only contracts with maturity so that age + maturity $\leq 110$. If age + maturity $> 110$, NA will be shown in the table.}\label{tab:annuity_tab_int} 
\begin{tabular}{@{}lrrrrrr@{}}
\toprule
 Age & $T=5$ & $T=10$ & $T=15$ & $T=20$ & $T=25$ & $T=30$ \\ 
\midrule
\underline{CPV} & \\
  60 & (4.497, 4.571) & (8.270, 8.495) & (11.363, 11.848) & (13.828, 14.678) & (15.722, 17.012) & (16.977, 18.800) \\ 
  65 & (4.454, 4.569) & (8.115, 8.486) & (11.060, 11.813) & (13.276, 14.542) & (14.746, 16.650) & (15.498, 18.021) \\ 
  70 & (4.372, 4.563) & (7.865, 8.455) & (10.488, 11.669) & (12.234, 14.154) & (13.114, 15.736) & (13.423, 16.475) \\ 
  75 & (4.218, 4.552) & (7.400, 8.359) & (9.541, 11.328) & (10.591, 13.265) & (10.975, 14.076) & (11.041, 14.274) \\ 
  80 & (3.978, 4.500) & (6.601, 8.050) & (8.002, 10.362) & (8.474, 11.375) & (8.546, 11.579) & (8.551, 11.598) \\ 
  85 & (3.521, 4.354) & (5.295, 7.252) & (5.894, 8.569) & (5.992, 8.856) & (6.004, 8.881)  & NA  \\ 
  90 & (2.888, 3.919) & (3.888, 5.719) & (4.047, 6.128) & (4.058, 6.162) &  NA & NA  \\ 
  95 & (2.204, 3.267) & (2.582, 4.022) & (2.610, 4.082) & NA   & NA  & NA  \\ 
  100 & (1.491, 2.258) & (1.598, 2.460) & NA   & NA  & NA  & NA  \\ 
  105 & (0.898, 1.561) & NA  & NA  & NA  & NA  & NA  \\ 
  \\
\underline{$L=6$} & \\
60 & (4.499, 4.572) & (8.286, 8.501) & (11.455, 11.864) & (14.033, 14.712) & (16.057, 17.097) & (17.460, 18.992) \\ 
  65 & (4.467, 4.571) & (8.174, 8.495) & (11.221, 11.839) & (13.547, 14.627) & (15.206, 16.825) & (16.164, 18.379) \\ 
  70 & (4.397, 4.569) & (7.983, 8.477) & (10.763, 11.768) & (12.699, 14.411) & (13.809, 16.234) & (14.281, 17.154) \\ 
  75 & (4.293, 4.563) & (7.596, 8.427) & (9.946, 11.544) & (11.258, 13.729) & (11.775, 14.881) & (11.892, 15.186) \\ 
  80 & (4.088, 4.537) & (6.953, 8.250) & (8.616, 10.851) & (9.271, 12.240) & (9.380, 12.659) & (9.389, 12.704) \\ 
  85 & (3.736, 4.453) & (5.873, 7.673) & (6.712, 9.376) & (6.897, 9.928) & (6.913, 9.971) & NA  \\ 
  90 & (3.112, 4.150) & (4.322, 6.396) & (4.571, 7.038) & (4.592, 7.113) & NA  & NA  \\ 
  95 & (2.370, 3.523) & (2.901, 4.571) & (2.942, 4.694) & NA   & NA  & NA  \\ 
  100 & (1.546, 2.507) & (1.682, 2.821) & NA   & NA   & NA   & NA   \\ 
  105 & (0.964, 1.617) & NA & NA  & NA  & NA  & NA  \\
      \bottomrule
\end{tabular}
\end{table}

The forecast uncertainties become larger as maturities increase from $T=5$ to $T=30$ for a given age. The forecast uncertainties also increase as the initial ages when entering contracts increase from 60 to 105 for a given maturity.

\section{Conclusion}\label{sec:7}

We proposed an adaptation of the \citeauthor{HU07}'s \citeyearpar{HU07} method to a CoDa framework. Using the Australian age-specific life-table death counts, we evaluate and compare the point and interval forecast accuracies among the CoDa, HU, LC and two na\"{i}ve RW methods for forecasting the age distribution of death counts. Based on the MAPE and mean interval score, the CoDa method with the ETS forecasting method and $L=6$ is recommended, as it outperforms the HU method, LC method and two na\"{i}ve RW and RWD for forecasting the age distribution of death counts. The superiority of the CoDa method is driven by the use of singular value decomposition to model the age distribution of the transformed death counts and the summability \textit{constraint} of the age distribution of death counts.

We apply the CoDa method to forecast age-specific life-table death counts from 2015 to 2064. We then calculate the cumulative survival probability and obtain temporary annuity prices. As expected, we find that the cumulative survival probability has a pronounced impact on annuity prices. Although annuity prices for an individual contract may be small, mispricing could have a dramatic effect on a portfolio, mainly when the yearly benefit is a great deal larger than one Australia dollar per year. 

There are a few ways in which this paper could be extended, and we briefly discuss four. First, a robust CoDa method proposed by \cite{FHR09} may be utilised, in the presence of outlying years. Second, the methodology can be applied to calculate other types of annuity prices, such as the whole-life immediate annuity or deferred annuity. Thirdly, we can consider cohort life-table death counts for modelling a group of individuals. Finally, we may consider other density forecasting methods as in \cite{KMP+18}, such as the log quantile density transformation method.

\bibliographystyle{agsm}
\bibliography{CoDa}

\vspace{.1in}

\section*{Appendix: supplementary material}

\begin{description}
\item[Annuity price calculation for the Australian male data] In Section~\ref{sec:6}, we present the annuity price calculation for the Australian female data. Since the mortality rates are different between male and female, the estimate of annuity prices ought to be different. Here, we present the annuity price calculation for the Australian male data.
\item[R code and data files] The R code for implementing the compositional data analysis approach and for producing point and interval forecasts is provided, along with the Australian age- and sex-specific life-table death counts.
\end{description}

\begin{center}
\large \large Supplementary materials \\
\vspace{.1in}
``Forecasting age distribution of death counts: An application to annuity pricing" by Han Lin Shang and Steven Haberman
\end{center}

\noindent \textbf{S1. Annuity price calculation for the Australian male data}\vspace{.1in}

In Table~\ref{tab:annuity_tab}, to provide an example of the annuity calculations, we compute the best estimate of the annuity prices for different ages and maturities for a male policyholder residing in Australia. We assume a constant interest rate $\eta = 3\%$ and hence zero-coupon bond is given as $B(0, \tau) = \exp^{-\eta\tau}$. 

\begin{table}[!htbp]
\centering
\tabcolsep 0.27in
  \caption{\small Estimates of annuity prices with different ages and maturities ($T$) for a male policyholder residing in Australia. These estimates are based on forecast mortality rates from 2015 to 2064. We consider only contracts with maturity so that age + maturity $\leq 110$. If age + maturity $> 110$, NA will be shown in the table.}\label{tab:annuity_tab} 
{\renewcommand{\arraystretch}{1}
\begin{tabular}{@{}lrrrrrr@{}}
\toprule
 Age & $T=5$ & $T=10$ & $T=15$ & $T=20$ & $T=25$ & $T=30$ \\ 
\midrule
\underline{CPV} & \\
60 & 4.4821 & 8.1694 & 11.1358 & 13.4387 & 15.0911 & 16.1213 \\ 
  65 & 4.4417 & 8.0150 & 10.7890 & 12.7794 & 14.0204 & 14.5930 \\ 
  70 & 4.3779 & 7.7766 & 10.2151 & 11.7356 & 12.4372 & 12.6353 \\ 
  75 & 4.2734 & 7.3396 & 9.2514 & 10.1336 & 10.3827 & 10.4173 \\ 
  80 & 4.0425 & 6.5631 & 7.7262 & 8.0546 & 8.1001 & 8.1027 \\ 
  85 & 3.6839 & 5.3838 & 5.8637 & 5.9303 & 5.9341 & NA \\ 
  90 & 3.0140 & 3.8651 & 3.9832 & 3.9899 & NA & NA \\ 
  95 & 2.2655 & 2.5799 & 2.5978 & NA & NA & NA \\ 
  100 & 1.5593 & 1.6481 & NA & NA & NA & NA \\ 
  105 & 1.0617 & NA & NA & NA & NA & NA \\ 
\\
\underline{$K=6$} & \\
60 & 4.4851 & 8.1866 & 11.1898 & 13.5540 & 15.2963 & 16.4257 \\ 
  65 & 4.4520 & 8.0641 & 10.9077 & 13.0033 & 14.3617 & 15.0298 \\ 
  70 & 4.4041 & 7.8712 & 10.4262 & 12.0825 & 12.8970 & 13.1469 \\ 
  75 & 4.3124 & 7.4903 & 9.5503 & 10.5634 & 10.8742 & 10.9212 \\ 
  80 & 4.1171 & 6.7858 & 8.0983 & 8.5010 & 8.5618 & 8.5655 \\ 
  85 & 3.7732 & 5.6287 & 6.1980 & 6.2840 & 6.2893 & NA \\ 
  90 & 3.1297 & 4.0899 & 4.2350 & 4.2438 & NA & NA \\ 
  95 & 2.3655 & 2.7229 & 2.7448 & NA & NA & NA \\ 
  100 & 1.6179 & 1.7171 & NA & NA & NA & NA \\ 
  105 & 1.0895 & NA & NA & NA & NA & NA \\ 
 \bottomrule
\end{tabular}}
\end{table}

To measure forecast uncertainty, we construct the bootstrapped age-specific life-table death counts, derive the survival probabilities and calculate the corresponding annuities associated with different ages and maturities. Given that we consider ages from 60 to 110, we construct 50-steps-ahead bootstrap forecasts of age-specific life-table death counts. In Table~\ref{tab:annuity_tab_int}, we present the 95\% pointwise prediction intervals of annuities for different ages and maturities, where age + maturity $\leq 110$.

\begin{table}[!thbp]
\centering
\tabcolsep 0.01in
  \caption{\small 95\% pointwise prediction intervals of annuity prices with different ages and maturities ($T$) for a male policyholder residing in Australia. These estimates are based on forecast mortality rates from 2015 to 2064. We consider only contracts with maturity so that age + maturity $\leq 110$. If age + maturity $> 110$, NA will be shown in the table.}\label{tab:annuity_tab_int} 
{\renewcommand{\arraystretch}{1}
\begin{tabular}{@{}lrrrrrr@{}}
\toprule
 Age & $T=5$ & $T=10$ & $T=15$ & $T=20$ & $T=25$ & $T=30$ \\ 
\midrule\underline{CPV} & \\
  60 & (4.540, 4.572) & (8.384, 8.500) & (11.582, 11.867) & (14.188, 14.739) & (16.146, 17.136) & (17.437, 19.004) \\ 
  65 & (4.510, 4.570) & (8.263, 8.487) & (11.319, 11.828) & (13.624, 14.612) & (15.147, 16.786) & (15.866, 18.174) \\ 
  70 & (4.461, 4.563) & (8.083, 8.453) & (10.814, 11.698) & (12.610, 14.239) & (13.460, 15.845) & (13.711, 16.480) \\ 
  75 & (4.367, 4.547) & (7.656, 8.341) & (9.809, 11.306) & (10.845, 13.191) & (11.137, 13.959) & (11.179, 14.106) \\ 
  80 & (4.137, 4.480) & (6.825, 7.989) & (8.109, 10.208) & (8.457, 11.104) & (8.502, 11.278) & (8.506, 11.290) \\ 
  85 & (3.727, 4.301) & (5.500, 7.047) & (6.015, 8.135) & (6.088, 8.332) & (6.092, 8.347) & NA  \\ 
  90 & (2.959, 3.761) & (3.806, 5.258) & (3.930, 5.557) & (3.938, 5.575) & NA  &  NA \\ 
  95 & (2.157, 2.980) & (2.451, 3.523) & (2.465, 3.568) & NA  & NA & NA \\ 
  100 & (1.460, 2.000) & (1.538, 2.137) & NA  & NA & NA & NA \\ 
  105 & (0.811, 1.434) & NA  & NA & NA & NA & NA  \\ 
  \\
  \underline{$K=6$} & \\
  60 & (4.542, 4.573) & (8.396, 8.505) & (11.631, 11.882) & (14.302, 14.773) & (16.375, 17.216) & (17.828, 19.201) \\ 
  65 & (4.520, 4.572) & (8.319, 8.498) & (11.437, 11.858) & (13.862, 14.700) & (15.560, 17.008) & (16.512, 18.656) \\ 
  70 & (4.477, 4.568) & (8.166, 8.479) & (11.039, 11.785) & (13.042, 14.475) & (14.159, 16.395) & (14.592, 17.442) \\ 
  75 & (4.406, 4.560) & (7.835, 8.416) & (10.235, 11.547) & (11.567, 13.793) & (12.075, 15.025) & (12.198, 15.418) \\ 
  80 & (4.230, 4.518) & (7.192, 8.201) & (8.848, 10.857) & (9.446, 12.278) & (9.564, 12.697) & (9.573, 12.784) \\ 
  85 & (3.923, 4.426) & (6.083, 7.594) & (6.893, 9.353) & (7.082, 9.900) & (7.091, 9.948) & NA  \\ 
  90 & (3.275, 4.077) & (4.460, 6.292) & (4.689, 6.994) & (4.713, 7.097) & NA & NA \\ 
  95 & (2.423, 3.494) & (2.907, 4.582) & (2.955, 4.740) & NA & NA & NA \\ 
  100 & (1.611, 2.651) & (1.739, 3.006) & NA & NA & NA & NA \\ 
  105 & (1.055, 1.840) & NA & NA & NA & NA & NA \\ 
  \bottomrule
\end{tabular}}
\end{table}

\end{document}